\documentclass[12pt,a4paper]{article}
\usepackage[utf8]{inputenc}
\usepackage[english]{babel}
\usepackage{amsmath,bm}
\usepackage{mathtools}
 \usepackage{amsfonts,amssymb}
 \usepackage{natbib}
 \usepackage{wrapfig}
 \usepackage{booktabs}
 \usepackage{siunitx}
 \usepackage{caption}
\usepackage[all]{xy}
\usepackage{tikz}
\usepackage{tikz-cd}
\usepackage{rotating}
\usepackage{float}
\usetikzlibrary{matrix,calc} 
\usepackage{ytableau}
\usetikzlibrary{graphs}
\usetikzlibrary {positioning}
\usepackage{float}
\usetikzlibrary{decorations.shapes}
\usetikzlibrary {decorations.pathmorphing}
\usepackage{pgfplots}
\usepackage{pgfkeys}
\usepackage{xypic}
\usetikzlibrary{arrows.meta}
\usetikzlibrary{matrix,arrows}
\usepackage[colorlinks=true]{hyperref}
\hypersetup{citecolor=black}
\hypersetup{urlcolor=black}
\usepackage{subcaption,graphicx}

\usepackage[left=2cm,right=2cm,top=2cm,bottom=2cm]{geometry}
\usepackage{authblk}

\title{\Large{A Structural Model for Detecting Communities in Networks}}

\author{Alex Centeno\thanks{The first draft of this document dates from September 2021 and was titled "Social Networks and Exchange Economy". I thank Shuyang Sheng, Department of Economics (UCLA) for having the valuable willingness to review our work.}\\
		Facultad de Ciencias Básicas, Universidad Católica del Maule, Talca, Chile\\
	\texttt{acenteno@ucm.cl}
}

\newtheorem{thm}{Theorem}
\newtheorem{lemm}{Lemma}
\newtheorem{Asumm}{Assumption}
\newtheorem{definition}{Definition}
\newtheorem{prop}{Proposition}
\newtheorem{exam}{Example}

\newcommand{\diag}{\text{diag}}
\date{\nonumber}
\providecommand{\keywords}[1]
{
	\small	
	\textbf{Keywords:} #1
}
\newcommand{\vect}{\text{vec}}
\newtheorem{rmk}[thm]{Remark}

\begin{document}
	\maketitle	
	\begin{abstract}
The objective of this article is to propose an econometric model of social interaction to identify and analyze the response actions of a set of partially rational players embedded in subnetworks in the context of social interaction and learning. We characterize the formation of strategic networks as a static game of interactions with incomplete information, where players maximize their utility depending on the connections they establish and multiple interdependent actions that allow group-specific player parameters. It is challenging to apply this type of model to real-life scenarios for two reasons: the computation of the Bayesian Nash equilibrium is very demanding, and the identification of social influence requires the use of excluded variables that are often not available. Based on the theoretical proposal, we propose a set of simulation equations and discuss the identification of the social interaction effect using a multimodal autoregressive network.
	\end{abstract}

\section*{\large Introduction}

Social networks influence a variety of individual behavioral outcomes, including educational attainment \citep{Calvo2009}, employment and social mobility \citep{Calvo2004,Patacchini2012}, crime rate \citep{Calvo2004,Kling2005,Patacchini2012}, risk distribution \citep{Ambrus2014,Fafchamps2007}, inequality \citep{Calvo2004r, Calvo2007}, consumption levels \citep{Ambrus2014,Moretti2011}, technology adoption \citep{Bandiera2006,Conley2010}, dissemination of micro-finance \citep{Banerjee2013,Banerjee2015,Jackson2015} and others. The main objective of this paper is to put forward an idea and its development in the identification and estimation of models of social network formation using simultaneous equations
of social interaction with binary outcomes to characterize the interdependence of individual choices both between actions and between players.

Since social networks are often the result of individual decisions, understanding network formation at the group level is relevant for research on peer effects, or more generally, network effects, such as good public policy making and, above all, equipping policymakers with tools to deal with interconnected socio-economic systems. Of course, the idea that individuals are affected by their peers motivates policies that try to manipulate peer composition for better outcomes. Identifying peer effects is notoriously challenging because of the reflection problem \citep{Manski1993} as well as due to spurious peer effects originating from group level effects. Random group allocation may be one way to overcome these identification problems. With groups formed at random, a random-effects specification for group-level characteristics can be adopted. An alternative approach consists in postulating that, conditional on observed group-level characteristics, group-level effects can be viewed as randomly assigned. Regression control techniques based on observed group characteristics then lead to a similar random effects specification, but without the need to appeal to random group assignment. We propose estimators that can accommodate both scenarios.

We study a discrete choice model to propose an alternative in the quest to identify best-response actions in the framework of interaction and social learning models that players exercise when intra- and inter-network interactions are available to them. In addition, we present a network formation analysis of social interactions in a static game scenario with incomplete information and a finite number of partially rational players embedded in an undirected network,\footnote{In this paper, a network type will describe the local structure around an agent along with the characteristics of each agent in this local sub-network. Of course, the size of each type of network depends on the specification of preferences, spatial measures, and a maximum number of connections, among other factors. A more detailed exposition can be seen in \cite{De2018}.}  whose utility function is linear-quadratic and depends on the own intentions to act and on a finite number of parameters such as own efforts and synergy parameters \citep{Brock2001,Brock2002,Brock2007}.

This paper generalizes the single-activity social interaction model with discrete choices to a
simultaneous-equation model \citep{Ballester2006,Cohen2018,Kelejian2004,Huang2020}. To motivate the specification of the econometric model, we consider an incomplete information game where individuals interact in multiple actions through a network. Also, we characterize the sufficient condition for the existence of a unique Bayesian Nash Equilibrium (BNE) of the game, which in turn guarantee the coherency and completeness of the econometric model \citep{Tamer2003}. Therefore, the first task is to identify the best response actions that correspond to BNE equilibria. To do so, we assume that the local interaction effect is not uniform across players of the same type and reflects strategic substitutability or complementarity in efforts across all pairs of players.\footnote{For example, when work information flows through friendship ties, employment outcomes vary between agents identical to their location in the network and the location of those ties. \citep{Calvo2004}. \cite{Durlauf2004} provides a comprehensive survey of the theoretical and empirical literature on peer effects.} The network pattern of a given type and its local complementarities is captured in the network itself. This description allows us to identify the information and social interaction processes of agent-based modeling given the study of global and local externalities for a general pattern of interdependencies.\footnote{The model-based approach gives priority to actions with the highest expected utility. But such policy does not consider the effect of the agent’s behavior on the learning process and ignores the contribution of the agent's activity to the exploration of the opponent’s strategy. In essence, every action affects the interaction process in two ways: i) the effect on the expected reward according to the current knowledge held by the agent y ii) the effect on the acquired knowledge, and hence, on future rewards expected to be received due to better planning \citep{Carmel1999}.}

The second task is the development of the properties of a dyadic index of multi-modal network generation to obtain consistent estimators, which allow the researcher to make inferences about the asymptotic properties, understand the topological structure of the network, and, accordingly, make correct decisions in the presence of observed/unobserved heterogeneity.\footnote{As the name suggests dyadic models, provide a statistical framework centered on pairs of nodes \citep{Renyi1959}. While bi-modal networks have been studied and established \citep{Everett2013,Huang2020} as those where there are only two types of agents and the linkage, policies are determined by agents from different sub-networks. The main function of bi-modal networks is to measure the interaction relation between the responses associated with two different nodes.} Given the importance of choice and interaction mechanisms in the construction and intensity of networks, we develop an alternative to some aspects of social network theory, for example, those presented by \cite{Acemoglu2014,Bala1998,De2018,Jackson1996,Johnson2003} and \cite{Sarkar2019}, where the agents decide whether or not to form a bond but do not determine its intensity in the interaction. Model under the attractiveness of incomplete information, because the dependence between actions due to strategic interactions can be eliminated by conditioning on publicly observed signals, a strategy also exploited in the literature on estimating games but in different contexts \citep{De2018}.

  We study the formation of networks through subnetworks to build a tool to understand the mechanism of communication and information for decision making and, therefore, mechanisms of interaction within the same network, even if the players are partially rational, it also provides a formal characterization about to learning and social interaction with partially rational players. Of course, in situations of uncertainty, when players have partial information, observational learning is a crucial component of interaction. Among the many possible mechanisms by which players learn from each other, observational learning describes the process by which a player draws inferences about the information held by other players based on observation of their behavior \citep{Mueller2021}.

The contributions of this paper aim at developing a micro foundation concerning the formation of social networks and identification of peer effects in a context of multivariate actions. We characterize the decision-making process in multiple activities in an environment of social interaction and learning. The model we consider has two important features. One, and as is common in the social network literature \citep{Acemoglu2012,Calvo2015}, the players enjoy a utility that is a function only of the links they possess. On the other hand, our model allows a certain type of complementarity or substitution in the actions they perform according to a synergy parameter. This particularity allows a combination of options in a context of social interaction that is not restricted among the possible actions of the players. Also, the econometric model implicit in the best response function contributes information to the autoregressive simultaneous equations model introduced by \cite{Kelejian2004} and extended by \cite{Cohen2018} to allow and study network effects by capturing intra-network externalities. In addition, as a model of social interaction, \citep{Bramoulle2009,Lee2010}, the model in this paper includes the \textit{endogenous peer effects}, where a players choice may depend on the choices/actions of other players of the same type, the \textit{contextual effect} where a players choice/action may depend on the exogenous characteristics of his peers; and \textit{correlated effect} where players in the same network tend to behave similarly because they have similar unobserved characteristics and/or face similar social environments. Also, the model studies the usual \textit{simultaneity effect} that is endemic in simultaneous equation models \citep{Kelejian2004,Drukker2021} and is known as the Manski reflection effect \citep{Manski1993}. Finally, our model includes a new type of social interaction effect, the \textit{cross-player type-pair effec}t, where the actions of a player of a given type may depend on the actions of his or her ties in related activities. Based on \cite{Bramoulle2009} and \cite{Cohen2018}, we provide identification conditions for these social interactions and simultaneity effects based on the topology of the underlying networks.

Section 1 introduces the theorical model and presents a baseline set of maintained assumptions. Section 2 presents the econometric model and identification of structural parameters. Section 3 presents the joint maximum likelihood estimations, transitive structure and estimation of econometric model. Section 4 correspond to Appendix of main results.

\section*{\large Related Literature}

The theoretical literature on the formation of social networks has flourished  \citep{Acemoglu2011,Ballester2006,Calvo2015,De2015,Jackson2007}. In econometric studies on the identification and structural estimation of social network formation models, there is still room for researchers to \citep{Ballester2006,Cohen2018,De2018,Elhorst2012,Graham2017,Huang2020,Leung2020,Sheng2020}. The main issues of social and economic interaction networks are to establish the emergence of social learning and bounded rational learning \citep{Acemoglu2011,Mueller2021}, exchange of information \citep{Acemoglu2014,Leung2020} and, development and stability of structural groups within the network itself \citep{Sheng2020,Stadtfeld2020}.\footnote{A network $G$ is pairwise stable according to \cite{Jackson1996} if there is any link present
	in $G$ that is mutually beneficial and any missing link is detrimental to at least one of the parties involved. Formally,
		\[\forall i,j\in G,\quad U_{i}(G)\geq U_{i}(G_{-ij})\quad y\quad U_{j}(G)\geq U_{j}(G_{-ij})\]
		y
		\[\forall i,j\notin G,\quad U_{i}(G)>U_{i}(G_{+ij})\quad o\quad U_{j}(G)>U_{j}(G_{+ij}),\]
		where $ij\notin G$ means that the link between $i$ and $j$ does not belong to the network. The network $G_{-ij}$ is the net without the edge $ij$ and $G_{+ij}$ is the network G with the link between $i$ and $j$. A network is efficient if the \textit{network value}, represented as $v:\{g:g\subset G\}\to \mathbb{R}$ satisfies $v(g)\geq v(g')$ $\forall g'\subset G$.} This is due to several reasons. First, social groups and, in particular, economic agents, are characterized by high within-group connectivity and a lack of between-group connectedness \citep{Moody2003,Stark2013}. This limitation has a non-positive effect on the flow of information and economic interaction. For example, in Scale-free social and economic networks, phenomena such as homophily are not sufficient to find equilibria in diverse structural groups, because although this property is regenerative, it also decreases the marginal importance of an eventually valuable \citep{Fischer1982,Yavacs2014}.
Second, a large number of theoretical models of economic networks at the micro level do not explain their emergence and development, but rather assume that a fully formed and configured network already exists and go on to study its properties \citep{Bala2000,Ehrhardt2007,Jackson2003,Jackson1996,Reggiani2015}. 

  \cite{Baetz2015} attempts to provide a micro basis on which agents optimize their link formation decisions. Also, some remarkable dynamic and strategic dynamic network formation models using linear utility functions have been presented, in which they assume that the largest eigenvalue of the relevant network is bounded regardless of the size of the network \citep{Currarini2009,Ghiglino2012,Jackson2007}.
  Bi-modal networks have been extensively studied \citep{Borgatti1997,Everett2013,Huang2020,Liu2014}. Unlike mono-mode models \citep{Lee2010,Chen20131}, where all nodes belong to a single network type and are treated equally. The model presented in this paper supports multiple types of nodes and it will be assumed that the network correlation coefficient is different for each type of player network. In particular, for each player of a given type, its associated response is modeled as a weighted average response of its connected neighbors of the other type. As a result, different network autocorrelation coefficients are allowed for players of different types. Applications of the model have large practical values and implications. For example, how information is aggregated between players of different network types.\footnote{\cite{Acemoglu2014} developed a framework for the analysis of information exchange through communication
	and investigated its implications for information aggregation in large societies. Su model draws close attention to two main features of social learning: First, the
		timing of actions is often endogenous and it is determined by the trade-off between the cost of waiting and the benefit of becoming more informed over time about the underlying environment. Second, the communication network typically imposes constraints on  the rate at which an agent acquires information and plays an important role in whether	agents end up taking “good” actions. }

On the other hand, the difficulty of preserving convergence to social learning with partially rational agents increases even in those networks where most of the agents involved are not neighbors, however, neighbors of any node may be linked and can indirectly reach other agents by a limited number of indirect links, this type of networks are also called Small-world networks \citep{Watts1998}. Also, in those networks where a finite number of links are assigned ''randomly'' to each agent, also called Ramdon networks \citep{Newman2002,Renyi1959} or, in those networks where the degree distribution follows the power-law asymptotically, also known as Scale-free networks \citep{Barthelemy2011}. These points require us to focus on two theoretical components of observational learning: Bounded Rational Bayesian and Network Learning.

In the Bayesian approach, agents are assumed to learn rationally, i.e., they make inferences about the private information of all agents based on the interaction structure and observed decisions \citep{Mueller2021}. This is the standard approach in the literature on sequential social learning \citep{Acemoglu2011,Arieli2017,Arieli2019,Bala1998} and in parts of the literature on repeated interaction in social networks \citep{Gale2003,Rosenberg2009,Mossel2015}. Although a useful benchmark, the Bayesian approach has a major weakness and it arises from the rationality assumption which seems unrealistic due to the computational sophistication needed to make inferences. This is especially true in an incomplete network where agents interact repeatedly. To reduce the cognitive complexity inherent in Bayesian updating, the Network Learning approach assumes instead that agents use simple rules. For this reason, the network learning approach is particularly employed, in complex environments, such as repeated interaction environments in social networks \citep{Mueller2021}.\footnote{\cite{Degroot1974} provides the standard model within the bounded rational approach. Its original formulation describes agents who repeatedly communicate beliefs about an underlying state of the world and revise their beliefs to a weighted average of their previous beliefs and those of their neighbors. A more recent formulation, the so-called DeGroot action model, is applied to the study of observational learning in environments with binary states and binary actions where agents, rather than communicating beliefs, observe actions \citep{Chandrasekhar2020}. As a bounded rational model, the DeGroot's model reduces the cognitive complexity of the updating process and its main strength derives from its tractability. However, this comes at the cost of a lack of generality, since the use of a weighted average update function in the crucial step of belief formation is somewhat arbitrary. Thus, both the Bayesian approach and the DeGroot model, as the standard formulation of the bounded rational approach, have weaknesses that limit their scope. This article is motivated by such unresolved weaknesses.} This paper aims to combine both alternatives, bounded rational Bayesian and network learning, to relate Bayesian updating, Bayesian Nash Equilibrium, and Social Interactions with Bounded Rationality and Incomplete Information.

 From an econometric point of view, to analyze network formation models with bounded degree, this paper develops some of the lines of future work and, also, uses some tools exposed by \cite{Graham2017} about two estimators for the homophily parameter. The first, tetrad logit (TL), estimator conditions on a
 sufficient statistic for the degree of heterogeneity. The second, joint maximum likelihood (JML), estimator treats the degree of heterogeneity  as additional (incidental) parameters to be estimated. The difference with our work is that we append the parameter of interdependencies that obeys some kind of transitivity in the structure of game networks \citep{De2018,De2020}. While \cite{Dzemski2019} study a dyadic linking model in which agents form directed links that exhibit
 homophily and reciprocity and consider specification testing and inference
 with respect to the homophily and reciprocity parameters. Also, the transitivity test can be interpreted as testing the dyadic model against models that
 target the formation of transitive relationships. This includes models of strategic network
 formation with agents who value transitive closure \citep{Leung2015,Mele2017}.
 
 Another problem in identifying network formation models with strategic interactions is the presence of multiple equilibria. Computing the complete set of equilibria is a generally difficult problem. An amateur search that checks the equilibrium conditions of each agent is computationally infeasible because the number of action profiles is exponential in the number of players. \cite{Fowler2010} circumvented the multi-policity problem by considering a sequential model that normally produces a single network or a single stationary distribution over networks such as those presented by \cite{Jackson2002e}. In comparison with these works and due to the specified utility, our model admits existence and uniqueness in the Bayesian Nash Equilibrium \citep{Liu2014,Liu2019}. Specifically, a single vector of best response functions throughout the network. In a way, we are homogenizing the network, by considering that all players have the same utility. However, an even more realistic way is to study utility types according to network types.

\begin{center}{\textbf{Notation}}
\end{center}
In the following, matrices will be denoted by bold capital letters and vectors by bold small letters. If \textbf{B} is a $n\times n$ matrix with $(i,j)$th element $b_{ij}$, then $\|\textbf{B}\|_{\max}=\sup_{i,j}|b_{ij}|$, $\|\textbf{B}\|_{\infty}=\sup_{i}\sum_{j=1}^{N}|b_{ij}|$. The vectorized of \textbf{B}, denoted $\vect(\textbf{B})=(b_{11},b_{21},\ldots,b_{N1},b_{12},\ldots, b_{NN})'$, $\textbf{F}(\cdot)$ will denote a distribution function with density function $f(\cdot)$.

\section*{\large Theorical Model and Baseline Assumptions}

In this section, we develop the network formation model. We assume a finite set of players $[n]=\{1,2,\ldots,n\}$ connected in some way in an undirected network $G\in\mathcal{G}$ with $G:n\times n\to\{0,1\}$. Each player $i$ has a set of observed attributes $X_{i}$ and a vector $\nu_{i}=(\nu_{i1},\ldots,\nu_{i,i-1},0,\nu_{i,i+1},\ldots,\nu_{i,n})$ of unobserved preferences, where $\nu_{ij}$ is the $i$-th preference for the link $ij$. To each pair of players $i$,$j$, corresponds to a degree of heterogeneity, $A_{ij}$. Let $X=(X'_{1},\ldots,X'_{n})'$ and $\nu=(\nu'_{1},\ldots, \nu'_{n})'$ an $(n-1)\times 1$ vector observed and unobserved respectively. For notation, $X_{ij,n}=(X_{i,n},X_{j,n})$  and $X_{n}=(X_{ij,n})_{i\neq j}$.

\subsection*{\small Link formation}
  A link represents an undirected relationship between two players. These links form a networks, which is denoted by $G\in \mathcal{G}$. It has a binary matrix representation, \textbf{G} of dimension $n\times n$, where $G_{ij}=1$ if players $i$ and $j$ are linked and $G_{ij}=0$ else. In particular, for all potential link $ij$, with $i\neq j$ we observe a dummy variable $G_{ij}$. In particular, the link rules \textit{direct} and \textit{indirect} are respectively:
\begin{align*}
G^{\wedge}_{ij}=&\textbf{1}\biggl\{
(i,j):X_{ij}'\beta+A_{ij}-\nu_{ij}\geq0\biggr\}\\ G^{\vee}_{ij}=&\textbf{1}\biggl\{(h,i,j):G^{\text{dir}}_{ih}= G_{jh}^{\text{dir}}=1\,\, \text{and}\,\, d(i,j)<\min\{ d(i,h),d(j,h)\}\biggr\}
\end{align*}

where \textbf{1}$(\cdot)$ denotes the indicator function, $d(i,j)$ is the distance between the player $i$ and $j$.\footnote{Following the terminology of \cite{Acemoglu2014}, a \textit{path} between player $i$ and $j$ in $G$ is a sequence $i_{1},i_{2},\ldots i_{K}$ of different players such that $i_{1}=i$, $i_{K}=j$, y $\{i_{k},i_{k+1}\}\in \mathcal{G}$ for $k\in\{1,\ldots, K-1\}$. The \textit{length} of a path is defined as $K-1$. The \textit{distance} between players $i$ and $j$ is:
	$d_{ij}=\min\{\text{length of}\, \mathcal{P}: \mathcal{P}\, \text{is a path from}\, i\, \text{to}\, j\, \text{on}\, \mathcal{G}\}$. In particular, the \textit{$\delta$-step} \textit{neighborhood} of player $i$ is defined as:
	$\mathbf{B}_{\delta}(i)=\{j: d_{ij}\leq \delta\}$.} The expression associated to $G^{\text{ind}}_{ij}$ refers to the non-remote connections in common that the players possess $i$ and $j$ \citep{De2020}. Let us denote the set where $G^{\vee}_{ij}$ is realized as $J$ a set compact of $\mathbb{R}^{|J|}$ and $|J|<n$. The interdependence component appeals to transitivity in network structures and is predicted by models of strategic network formation through preference diffusion \citep{De2018}.\footnote{Transitivity in ties may arise for two distinct reasons. First, agents may have a structural taste for transitive links. The benefits of link formation between any two agents may cause the number of neighbors they have in common to increase. \cite{Coleman1994} provides an impetus for the agents to form transitive relation. \cite{Jackson2011} provides a game-theoretic basis for transitivity, arguing that common friends, by supervising transactions between agents, help maintain cooperation. So, socialization may be easier and more enjoyable when individuals share common friends. Second, transitivity in social networks may reflect assortative matching on an unobserved attribute. That is, the benefits of bonding may be greater among similar agents, leading to dense ties among them.}
The expression $X'_{ij}$ is a homophily function and varies by pairs of players concerning for to the observed attributes \citep{Currarini2009}. 
 The terms $\{A_{ij}=A_{i}+A_{j}\}_{(i,j)\in n^{2}}$,  vary with respect to the degree of unobserved heterogeneity of the players $i,j$ and, finally an idiosyncratic component, $\nu_{ij}$, assumed independent and identically distributed between pairs of players. If there is no room for confusion for the reader, we can compact the link rule between players $i$ and $j$  in the network as follows:
\begin{equation}\label{eq:Gij}
	G_{
		ij	}=\textbf{1}\biggl\{\big(\sum_{k\in J}G_{ik}G_{jk}\big)\delta+X_{ij}'\beta+A_{ij}-\nu_{ij}\geq0\biggr\}
\end{equation}
 Rule \eqref{eq:Gij} is more general instead of taking $\delta=0$, which would imply that only direct entailments are important \citep{Graham2017}. For any $i\neq j$, we descompose $G$ into $(G_{ij},G_{-ij})$, where $G_{-ij}\in\mathcal{G}_{-ij}$ is the network obtained from $G$ by removing link $ij$. 
 
 An players degree equals the number of links she has: $G_{i+}=\sum_{i\neq j}G_{ij}$. The row (or column) sum of the adjacency matrix, denoted by the $n\times 1$ vector $\textbf{G}_{+}=(G_{1+},\ldots, G_{n+})'$, give the network's \textit{degree sequence}.
 
 
  We consider the following data generating process. The players are divided on $h$ types (neighborhoods). Without restriction, links can exist between players of different types.
 The variable $h$ will denote the players of type $h$. Let $n_{h}$ be an integer generated from a distribution on $\{2,3,\ldots\}$ denoting the number of players of type $h$ such that $\sum_{h}n_{h}=n$. Each player $i$ of type $h$ is associated with a vector of observed attributes $X_{i,h}$ and a vector of
 unobserved preferences $\epsilon_{i,h}$. We define the complement $G^{c}_{h}=\{(i,j):i, j\notin n_{h}, i\neq j\}$. Let $X_{ij,h} = (X_{i,h}, X_{j,h})$ the attributes of the pair of players $(i,j)$ and $X_{h} = (X_{ij,h})_{i\neq j}$ the attribute profile of all the pairs. We
 observe $h$ independent networks and their attribute profiles $(G_{h}, X_{\tau_{p}})_{h}$. The adjacency matrix corresponding to the super-network $G$ is a block matrix $\textbf{G}=\diag[\textbf{G}_{h}]_{h}$ with $G_{ij,h}=\textbf{1}\big\{\{G_{ij}=1\}\cap \big\{j\in G_{h}\big\}\big\}$. Therefore, our model directly relates the model specification and parameter space definition in higher order spatial econometric models and interaction models in social networks through subnetworks \citep{Elhorst2012,Leung2020}.
 \begin{figure}[t!]
 		\centering
 	\includegraphics[scale=0.7]{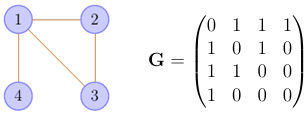}
 		\caption{Tailed Triangle Network.}

\end{figure}
\subsection*{\small Utility}

 Each player $i$ in the network has $r$ distinct ways of acting, collected in a vector $\textbf{a}_{i}=(a_{i1},a_{i2},\ldots,a_{ir})\in\{0,1\}^{r}$. Coming from Learning in Networks, each choice $a_{ik}$ corresponding to player $i$ in action  $k$ depends on the actions of its nearest neighbors. We try to model this dependence by the expression $\mathbb{P}(a_{ik}=1|\bigcup\{j:j\in \textbf{B}_{\delta}(i)\}:a_{jk}=1, k=1,\ldots, r)$. Following to \cite{Austin2013} and \cite{Lesage2009} for exogenous predictors $\overline{\textbf{x}}_{ik}=(\overline{x}_{1k},\ldots,\overline{x}_{(-i)k},\ldots,\overline{x}_{nk})$ and noises corresponding to actions $\mu_{ik}=(\mu_{ik},\ldots,\mu_{(-i)k},\mu_{ik})$,  be
 \begin{equation}
 	y_{ik}=\sum_{j\in \textbf{B}_{\delta}(i), j\neq i}G_{ij}a_{jk}+\overline{\textbf{x}}_{ik}\overline{\beta}-\mu_{ik}
 \end{equation}
 
 a latent continuous variable such that $a_{ik}=\textbf{1}\{y_{ik}>0\}$. The model can be further written as a Cliff-Ord type spatial model \citep{Kelejian2004}. We interpret $y_{ik}$ as the underlying intention (a priori knowledge) and $a_{ik}$ as the actual outcome choice \citep{Maddala1986}. Naturally, player $i$ will know his intentions $y_{i1},\ldots,y_{ir}$  according to the intentions of his links.
 The utility of player $i$ in the network depends explicitly on the network configuration and on an i.i.d component $\epsilon_{i}=(\epsilon_{i1},\ldots,0\ldots,\epsilon_{ir})'$ over each player $i$ and independent of the collection $w_{ik}$ $(k=1,\ldots, r)$ corresponds to the weighting associated with player $i$'s actions.
 
 Each type of player $g$ corresponds to a subnetwork $G_{g}$ (semi-dense)\footnote{In this article, a network is called semi-dense if for every pair of players $i, j$ of type $h$ it is satisfied that $G_{ij}=0$ except for a finite number of players of the same type \citep{Alizadeh2017}. Strictly speaking, if $\mu$ is a Borelian measure and $G$ an undirected lattice embedded in $\mathbb{R}^{d}$ for some sufficiently large $d$. It is said that $\mathcal{G}$ is semi-dense if $\mu\bigg(\{j\in G:G_{ij}=0\}\bigg)=0$ \citep{Renyi1959}.} and with itself a \textit{socio-matrix} binary symmetric $\textbf{G}_{g}$ of size $n_{g}\times n_{g}$, with elements $G_{ij,g}$. Also, this socio-matrix is fully determined by the set of pairs of players $\{ij\}_{(i,j)\in G_{g}}$, and is called \textit{the relative adjacency socio-matrix} to network $G_{g}$.  Let $G^{*}_{ij,g}=G_{ij,g}/n_{g}$ denote the corresponding normalized weights, let $\overline{a}_{il,g}=\sum_{j=1}^{n}G^{*}_{ij,g}a_{jl}$ denote the average actual outcome choice for action $k$ by the $g$-group of peer. The utility of player $i$ is: 
 \begin{multline}\label{eq:Utility}	U_{i}(G,\textbf{a})=\underbrace{\sum_{k,l=1}^{r}\sum_{g=1}^{h}\binom{n}{g }\binom{n_{g}}{n_{g}-(1-a_{ik})}\mathbb{P}(i\in G_{g}|\text{card}\{j:G_{ij,g}=1\})\big(s_{lk,g}\overline{a}_{il,g}+w_{ik,g}-\epsilon_{ik,g}\big)y_{ik,g}}_{\text{payoff}}-\\-\underbrace{\dfrac{1}{2}\sum_{k,l=1}^{r}\sum_{g=1}^{h}\varphi_{lk,g}y_{ik,g}y_{il,g}}_{\text{cost}}
 \end{multline}
 The the marginal utility of $i$ from forming a link with $j$ is:
 \begin{equation}\label{eq:esp}
 	\Delta(U_{ij}(G_{-ij},\textbf{a}))=\dfrac{1}{n_{g}}\sum_{k,l=1}^{r}\binom{n}{g }\binom{n_{g}}{n_{g}-(1-a_{ik})}s_{lk,g}a_{jl,g}y_{ik,g}
 \end{equation}
 for some subnetwork $g$ and it consists of the strength of synergy represented by both players in performing a set of actions for the expected value of performing a given action and the underlying intent of that action. The utility \eqref{eq:Utility} consists of two components: payoffs and cost. The parameter $s_{lk}$ ($l,k=1,\ldots,r$) captures the strategic complementarity or substitutability (its sign depends) between player $i$'s own effort in action $l$ and the average effort of his ties in action $k$, i.e., $s_{lk}>0$, an increase in effort in performing action $l$ produces a positive change in the effort to perform action $k$ \citep{Ballester2006}. We say that the efforts to perform actions $k$ and $l$ are strategic from the perspective of acting $k$. Reciprocally, when $s_{lk}<0$, these efforts are strategic substitutes from the perspective of acting $k$. The cost of player $i$'s effort in action $k$ depends on player $i$'s effort in all actions. The parameter $\varphi_{kl}$ measures the substitutability or complementarity (depending on its sign) of the effort levels in all actions of player $i$.\footnote{In the terminology of \cite{Manski1993}, the magnitude of $s_{lk}$ is an index of synergy in the endogenous social interaction between the effort expended by player $i$ for action $k$ and the effort expended by his links in action $l$.}
This utility differs from \cite{Cohen2018} and \cite{Liu2019}. First, because $y_{ik}$ represents the unobservable intention, corresponding to learning in networks and the utility of player $i$ depends on the choice of actions of its binders, while, in \cite{Cohen2018}, $y_{ik}$ represents the observable action on a set of continuous actions and the utility of player $i$ depends directly on $\{y_{jk}\}_{k=1}^{r}$ for all $j\in G$ with $G_{ij}=1$. Both assume $h=1$, i.e, only group for peer effects and they also do not consider the number of networks that can be formed, in addition to the number of players in each network that perform the same action. In \cite{Liu2019}, $y_{ik}$ represents a latent continuous variable not modeled by learning in networks. Second, \eqref{eq:Utility} captures network externalities via the likelihood ratio $G_{ij}$.  \cite{Cohen2018} and \cite{Liu2019} do not assume the linking conditions between pairs of players in the network. As a result, maximization of the utility \eqref{eq:Utility} motivates a discrete-choice simultaneous-equation econometric model of social interactions, while the network game in \cite{Cohen2018} leads to a linear simultaneous-equation econometric model of social interactions.
 
\subsection*{\small Equilibrium}
 
 We utilize a refinement criterion named “local stability” on BNE. Local stability implies that the game converges to the original equilibrium if there is an infinitesimal perturbation in one or more players equilibrium behaviors. Using such a refinement scheme has important theoretical and practical implications. First, if an equilibrium is not stable, a slight deviation by one of the consumers will cause other player to deviate further away from the equilibrium. In such situations, it is, thus, practically difficult, if not impossible, for the system to reach the equilibrium. Second, an important advantage of structural models is its ability to perform counterfactual analysis. Without the equilibrium refinement, we need to compute all possible equilibria, which might lead to multiple and possibly contradicting policy implications.
 
 Given the network topology and the observable components of the productivity, the players simultaneously choose $y_{ik}$, for $k = 1,\ldots, r$, to maximize their expected utilities:
 \begin{multline}\label{eq:Utility_esp}	\mathbb{E}(U_{i}(G,\textbf{a}))=\sum_{k,l=1}^{r}\sum_{g=1}^{h}\binom{n}{g }\binom{n_{g}}{n_{g}-(1-a_{ik})}\mathbb{P}(i\in G_{g}|\text{card}\{j:G_{ij,g}=1\})\big(s_{lk,g}\psi_{jl,g}+w_{ik,g}-\epsilon_{ik,g}\big)y_{ik,g}-\\-\dfrac{1}{2}\sum_{k,l=1}^{r}\sum_{g=1}^{h}\varphi_{lk,g}y_{ik,g}y_{il,g}
 \end{multline}

 where $\psi_{jl,g}=\mathbb{E}_{w}(\overline{a}_{jl,g})$.
 
 From the first order condition of utility maximization, we have:
 \begin{equation}\label{eq_first}
 	y^{*}_{im,\tau_{p}}=\sum_{l=1,l\neq m}^{r}\phi_{lm,\tau_{p}}y^{*}_{il,\tau_{p}}+\sum_{l=1}^{r}\sum_{\tau_{p}=1}^{T}\lambda_{lk,\tau_{p}}\overline{a}_{il,\tau_{p}}+\overline{w}_{im,\tau_{p}}-\overline{\epsilon}_{im,\tau_{p}}
 \end{equation}
 with $\phi_{lk,\tau_{p}}=-\varphi_{lk,\tau_{p}}/\varphi_{mm,\tau_{p}}$, $\lambda_{lk,\tau_{p}}=s_{lk,\tau_{p}}/\varphi_{mm}$, $\overline{w}_{im,\tau_{p}}=w_{im,\tau_{p}}/\varphi_{mm,\tau_{p}}$ y $\overline{\epsilon}_{im,\tau_{p}}=\epsilon_{im,\tau_{p}}/\varphi_{mm,\tau_{p}}$. In matrix form, \eqref{eq_first} can be written as:
 \begin{equation*}\label{eq:U2}
 	\textbf{y}^{*}_{m,\tau_{p}}=\sum_{l=1, l\neq m}^{r}\phi_{lm,\tau_{p}}\textbf{y}^{*}_{l,\tau_{p}}+\sum_{l=1}^{r}\sum_{\tau_{p}=1}^{T}\lambda_{lm,\tau_{p}}\textbf{G}_{\tau_{p}}\bm{\psi}_{l,\tau_{p}}+\overline{\textbf{w}}_{m,\tau_{p}}-\overline{\bm{\varepsilon}}_{m,\tau_{p}}
 \end{equation*}
 Let \[\textbf{Y}^{*}_{\tau_{p}}=[\textbf{y}^{*}_{1,\tau_{p}},\textbf{y}^{*}_{2,\tau_{p}},\ldots, \textbf{y}^{*}_{r,\tau_{p}}],\quad \bm{\Psi}_{\tau_{p}}=[\bm{\psi}'_{1,\tau_{p}},\bm{\psi}_{2,\tau_{p}},\ldots,\bm{\psi}'_{r,\tau_{p}}]'\quad \textbf{W}=[\overline{\textbf{w}}_{1,\tau_{p}}, \overline{\textbf{w}}_{2,\tau_{p}},\ldots, \overline{\textbf{w}}_{r,\tau_{p}}]\quad \text{and}\]
 \[\bm{\epsilon}_{\tau_{p}}=(\overline{\bm{\varepsilon}}_{1,\tau_{p}},\overline{\bm{\varepsilon}}_{2,\tau_{p}},\ldots, \overline{\bm{\varepsilon}}_{r,\tau_{p}})\] Then
 \begin{equation}\label{eq:Matricial}
 	\textbf{Y}^{*}_{\tau_{p}}	=\textbf{Y}^{*}_{\tau_{p}}\bm{\Phi}_{\tau_{p}}+\textbf{G}_{\tau_{p}}\bm{\Psi}_{\tau_{p}}\bm{\Lambda}_{\tau_{p}} + \textbf{W}_{\tau_{p}}-\bm{\epsilon}_{\tau_{p}}
 \end{equation}
 where $\bm{\Phi}_{\tau_{p}}=[\phi_{lk,\tau_{p}}]$ and $\bm{\Lambda}_{\tau_{p}}=[\lambda_{lk,\tau_{p}}]$ are parameter matrices $r\times r$. The off-diagonal elements of $\bm{\Phi}_{\tau_{p}}$, $\phi_{lk,\tau_{p}}$, represent the \textit{effect of simultaneity}, i.e., an individual choice in an action $k$ may depend on his own choices in action $l$. The diagonal element of $\bm{\Lambda}_{\tau_{p}}$, $\lambda_{kk,\tau_{p}}$, represents the \textit{within-activity peer effect}, where an players choice in an action $k$ may depend on the expected choices of the peers in the same activity. The off-diagonal element of $\bm{\Lambda}_{\tau_{p}}$, $\lambda_{lk,\tau_{p}}$, represents the \textit{cross-activity peer effect}, where an players choice in an action $k$ may depend on the expected choices of the peers in a related activity $l$.
 If $\textbf{I}_{r,\tau_{p}}-\bm{\Phi}_{\tau_{p}}$ is non-singular, then the reduced form of the model's \eqref{eq:Matricial} is:
 \begin{equation}\label{eq:reduced}
 	\textbf{Y}^{*}_{\tau_{p}}=\textbf{G}_{\tau_{p}}\bm{\Psi}_{\tau_{p}}\tilde{\bm{\Lambda}}_{\tau_{p}}+\tilde{\textbf{W}}_{\tau_{p}}-\tilde{\bm{\epsilon}}_{\tau_{p}}
 \end{equation}
 with $\tilde{\bm{\Lambda}}_{\tau_{p}}=\bm{\Lambda}_{\tau_{p}}(\textbf{I}_{r}-\bm{\Phi}_{\tau_{p}})^{-1}$, $\tilde{\textbf{W}}_{\tau_{p}}=\textbf{W}_{\tau_{p}}(\textbf{I}_{r}-\bm{\Phi}_{\tau_{p}})^{-1}$ and $\tilde{\bm{\epsilon}}_{\tau_{p}}=\bm{\epsilon}_{\tau_{p}}(\textbf{I}_{r,\tau_{p}}-\bm{\Phi}_{\tau_{p}})^{-1}$. From \eqref{eq:reduced},
 \begin{equation*}
 	y_{im,\tau^{*}_{p}}=\sum_{l=1}^{r}\sum_{\tau_{p}=1}^{T}\tilde{\lambda}_{lm,\tau_{p}}\sum_{j=1}^{n}G_{ij,\tau_{p}}\psi_{jl,\tau_{p}}+\tilde{w}_{im,,\tau_{p}}-\tilde{\epsilon}_{im,,\tau_{p}}
 \end{equation*}
 where $\tilde{\lambda}_{lm,,\tau_{p}}$, $\tilde{w}_{im,,\tau_{p}}$ y $\tilde{\epsilon}_{im,,\tau_{p}}$ are the elements $(l,m)$ and $(i,m)$ de $\tilde{\bm{\Lambda}}_{\tau_{p}}$ y $\tilde{\textbf{W}}_{\tau_{p}}$ y $\tilde{\bm{\epsilon}}$ respectively. Then,
 \begin{equation*}
 	\mathbb{P}_{w}(a_{im,\tau_{p}}=1|R_{[j]})=\textbf{F}_{k,\tau_{p}}(\sum_{l=1}^{r}\sum_{\tau_{p}=1}^{T}\tilde{\lambda}_{lm,\tau_{p}}\sum_{j=1}^{n}G_{ij,\tau_{p}}\psi_{jl}+\tilde{w}_{im,\tau_{p}})
 \end{equation*}
 with $R_{[j]}=\bigcup\{j:j\in \textbf{B}_{\delta}(i):a_{jm}=1\}$ and $\textbf{F}_{k,\tau_{p}}$ is a distribution function of $\tilde{\epsilon}_{im,\tau_{p}}$.
 
 Let $\textbf{y}_{\tau_{p}}=\vect(\textbf{Y}_{\tau_{p}}^{*})$, $\bm{\psi}_{\tau_{p}}=\vect(\bm{\Psi}_{\tau_{p}})$, $\textbf{w}_{\tau_{p}}=\vect(\tilde{\textbf{W}_{\tau_{p}}})$ and $\bm{\epsilon}_{\tau_{p}}=\vect(\tilde{\bm{\epsilon}}_{\tau_{p}})$. As in \cite{Liu2019}, be it
 \begin{equation}\label{eq:eq}
 	\textbf{g}(\bm{\psi}_{\tau_{p}})=[\textbf{g}_{1}(\bm{\psi}_{1})', \textbf{g}_{2}(\bm{\psi}_{2})', \ldots, \textbf{g}_{r}(\bm{\psi}_{T})']'
 \end{equation}
 where $\textbf{g}_{m}(\bm{\psi}_{\tau_{p}})=[\textbf{F}_{m}(q_{1m,\tau_{p}}), \textbf{F}_{m}(q_{2m,\tau_{p}}),\ldots, \textbf{F}_{k}(q_{nm,\tau_{p}})]'$ for all $m=1,\ldots, r$ with $$q_{im,\tau_{p}}=\sum_{l=1}^{r}\sum_{\tau_{p}=1}^{T}\tilde{\lambda}_{lm,\tau_{p}}\sum_{j=1}^{n}G_{ij,\tau_{p}}\psi_{jl,\tau_{p}}+\tilde{w}_{im,\tau_{p}}.$$ In the Bayesian Nash Equilibrium in $G_{\tau_{p}}$, $\bm{\psi}_{\tau_{p}}=\textbf{g}(\bm{\psi}_{\tau_{p}})$ \citep{Liu2019,Osborne1994}.
 \begin{rmk}\label{rmk}
 	If we define $U_{\tau_{p}}$ as the aggregate utility of the players type $\tau_{p}$ and $U_{\tau_{q}}$ as the aggregate utility of the players type $\tau_{q}$. Let us denote by $U_{\tau_{p},-\tau_{q}}$ as the aggregate utility of the players type $\tau_{p}$ without linking to players of type $\tau_{q}$ and $U_{\tau_{p},+\tau_{q}}$ the aggregate utility of type players $\tau_{p}$ by linking up with the network of players type $\tau_{q}$. Thus, we modeled the link of networks $G_{\tau_{p}}$ and $G_{\tau_{q}}$ according to $\mathbb{E}(U_{\tau_{p},+\tau_{q}}>U_{\tau_{p},-\tau_{q}}|U_{\tau_{p}}, U_{\tau_{q}})$. In particular, \[G_{\tau_{p}\tau_{q}}=\textbf{1}\biggl\{\bigcup_{(i,j)\in G_{\tau_{p}}\times G_{\tau_{q}}}\{ij\}:\mathbb{E}(U_{ij}(G_{ij}))>\mathbb{E}(U_{ji}(G_{-ij}))\biggr\}\]
 	where $\mathbb{E}(U_{ij}(G_{ij}))$ represents the expected utility of linking player $i$ to player $j$ and $\mathbb{E}(U_{ji}(G_{-ij}))$ represents the expected utility of player $j$ without linking to player $j$. Of course, linking with players of another type entails a greater effort reflected in the marginal cost of such a connection with respect to any pair of actions, respectively. In this case,
 	\begin{equation*}
 		\Pi_{i,\tau_{p}}(G_{\tau_{p}},\textbf{a})(g_{\tau_{q}})=U_{i,\tau_{p}}(G,\textbf{a})-\sum_{(i,j)\in g(t)}t^{i}_{ij,\tau_{q}}
 	\end{equation*} 
 	where $t^{i}_{ij,\tau_{q}}=\sum_{j\notin G_{\tau_{p}}}\mathbb{E}(U_{ij}(G_{-ij}),y^{*}_{ik}, R_{[j]})$, $g_{\tau_{q}}(t)=\{(i,j)\in  G_{\tau_{p}}\times G_{\tau_{q}}:t^{i}_{ij,\tau_{p}}+t_{ji,_{\tau_{q}}}^{j}\geq 0\}$. 
 
 	While \cite{Jackson1996}, \cite{Jackson2001}, \cite{Jackson2003} are reduced to the case $G_{\tau_{p}\tau_{q}}=0$ for all types $\tau_{p}\neq \tau_{q}$ i.e. all players in the network are treated as of equal type and, therefore, the existence of equilibria and pairwise stability is guaranteed \citep{Jackson1996}. Hence, the motivation to impose another type of linkage, strong ties, in the theory of social networks.\footnote{\cite{Cohen2018} emphasizes that the use of multiple reference groups makes it possible to distinguish between contextual and endogenous effects. Characteristics to which \cite{Manski1993} pointed the way to the inability to identify structural parameters. However, \cite{Cohen2018} addressed this ongoing challenge in the social interactions literature, in particular, the continued difficulty in discriminating between any type of social interaction and unobserved group-level heterogeneity.}
 \end{rmk}
 
Following the ideas \cite{Graham2017}, or the interdependence structure, let \textbf{X} be the $n\times \dim(X)$ matrix of observed player attributes and $\textbf{A}$ the $n×
\times 1$ vector of unobserved player-level degree heterogeneity terms. Some of the results presented below maintain the following two assumptions, with additional assumptions made for specific results.
\begin{Asumm}(Likelihood)
	The conditional likelihood of the network $\textbf{G}=\textbf{g}$ is
	\begin{equation}
		\mathbb{P}(G_{ij}=g|\textbf{X},\textbf{A})=\prod_{i< j}\mathbb{P}(G_{ij}=g|\textbf{X},\textbf{A})
	\end{equation}
with 
\begin{multline}
	\mathbb{P}(G_{ij}=g|\textbf{X},\textbf{A})=\Biggl[\dfrac{1}{1+\exp(\big(\sum_{k\in J}G_{ik}G_{jk}\big)\delta+X_{ij}'\beta+A_{ij})}\Biggr]^{1-g}\times\\\times\Biggl[\dfrac{\exp(\big(\sum_{k\in J}G_{ik}G_{jk}\big)\delta+X_{ij}'\beta+A_{ij})}{1+\exp(\big(\sum_{k\in J}G_{ik}G_{jk}\big)\delta+X_{ij}'\beta+A_{ij})}\Biggr]^{g} \forall i\neq j.
\end{multline}
\end{Asumm}
Assumption 1 implies that the idiosyncratic component of link surplus, $\nu_{ij}$, is a standard logistic random variable that is independently and identically distributed across of pairs of players. Assumption 1 also implies that links form independently conditional on \textbf{X} and \textbf{A}. Importantly, independence is conditional on the latent player attributes $\{A_{i}\}_{i=1}^{n}$ \citep{Graham2017}. This type of dependence
is analogous to that allowed for by a strict exogeneity assumption in a single agent static panel data model \citep{Chamberlain1984}. As shown by \cite{Graham2017}, unconditionally on these attributes, independence does not hold. The logistic assumption is important for the tetrad logit (TL) estimator, but less so for the joint maximum likelihood (JML) estimator. The assumption that links form independently of one another conditional on player attributes will be plausible in some settings, but not in others. In particular, rule \eqref{eq:Gij} and Assumption
1 are appropriate for settings where the drivers of link formation are predominately bilateral in nature, as may be true in some types of friendship and trade networks as well as in models of (some types of) conflict between nation-states \citep{Silva2006}. The approach taken here is to study identification and estimation issues when links
form according to rule \eqref{eq:Gij} and Assumption 1. This setting both covers a useful class of empirical examples and represents a natural starting point for formal econometric analysis. An analogy with single agent discrete choice panel data models is perhaps useful. In that
setting, early methodological work focused on introducing unobserved correlated heterogeneity into static models of choice\citep{Chamberlain1984}.

To characterize the large sample properties of the JML estimates, we require some additional notation and an identification condition. It is useful to begin by observing that the population problem is
\[\max_{(b_{1},b_{2})\in \mathbb{A}\times\mathbb{B},\textbf{o}_{n}\in \mathbb{I}^{n}}\mathbb{E}\biggl[L_{n}(b_{1},b_{2},\textbf{o}_{n}|\textbf{X}, \textbf{A}_{n_{0}})\biggr]\]
where it is easy to show that
\[\mathbb{E}\biggl[L_{n}(b_{1},b_{2},\textbf{o}_{n}|\textbf{X}, \textbf{A}_{n_{0}})\biggr]=-\sum_{i< j}D_{KL}(p_{ij}||p_{ij}(\theta,A_{i},A_{j}))-\sum_{i\neq j}\mathbb{H}(p_{ij})\]
with $\textbf{A}_{n}$ denote an $n \times1$ vector of degree heterogeneity values and $\textbf{A}_{n_{0}}$ the corresponding vector of true values, 
\[p_{ij}(\theta,\textbf{A})=\dfrac{\exp(\big(\sum_{k\in J}G_{ik}G_{jk}\big)\delta+X_{ij}'\beta+\iota'_{ij}\textbf{A})}{1+\exp\big(\big(\sum_{k\in J}G_{ik}G_{jk}\big)\delta+X_{ij}'\beta+\iota'_{ij}\textbf{A}\big)},\quad \theta=(\beta,\delta),\]
 $L_{n}(\theta,\textbf{A}_{n})=\sum_{i< j}\biggl\{G_{ij}\ln p_{ij}(\theta,\textbf{A})+(1-G_{ij})\ln [1-p_{ij}(\theta,\textbf{A})]\biggr\}$
 where $D_{KL}(p_{ij}||p_{ij}(\theta,A_{i},A_{j}))$ is the Kullback–Leibler divergence of $p_{ij}(\theta,A_{i},A_{j})$  from $p_{ij}:=p_{ij}(\theta_{0},A_{i_{0}},A_{j_{0}})$ and $\mathbb{H}(p_{ij})$ is the binary entropy function. It is clear that $(\theta_{0},A_{i_{0}},A_{j_{0}})$ is
a maximizer of the population criterion function. The following assumption ensures that it is the unique maximizer (and also that this maximizer exists for large enough $n$).
\begin{Asumm}(Joint Identification)
	For $i=1,\ldots,n$ the support of $A_{i_{0}}$ es $\mathbb{I}$, a compact subset de $\mathbb{R}$ and\[\mathbb{E}\big[L_{n}(b_{1},b_{2},\textbf{o}_{n}|\textbf{X}, \textbf{A}_{n_{0}})\big]\] is is uniquely maximized at $b_{1}=\beta_{0}$, $b_{2}=\delta_{0}$ and $\textbf{o}_{n}=\textbf{A}_{n_{0}}$ large enough $n$.
\end{Asumm}

	\begin{Asumm}
		\begin{itemize}
			\item[(i)](Regularity) $(\beta,\delta)\in \text{int}(\mathbb{A})\times \text{int}(\mathbb{B})$, with $\mathbb{A}\times \mathbb{B}$ a compact subset of $\mathbb{R}^{\dim(\beta)}\times\mathbb{R}^{\dim(\delta)}$. The component $X'_{ij}$ is a known transformation of $(X'_{i},X'_{j})'$. The support of $X'_{ij}$ is $\mathbb{X}$, a compact subset of $\mathbb{R}^{\dim(\beta)}$.
			\item[(ii)] (Random Sampling) Let $i\in \{1,\ldots,n\}$ index a random sample of players
			from a population. The econometrician observes $(G_{ij},X_{ij})$ for  $i\in \{1,\ldots,n\}$, $j\neq i$.
		\end{itemize}
\end{Asumm}
Part (i) of Assumption 3 is standard in the context of nonlinear estimation problems. It implies that the observed component of link surplus, $X_{ij}'\beta$ and $\big(\sum_{k\in J}G_{ik}G_{jk}\big)\delta$, will have bounded support. This simplifies the proofs of the main theorems, especially those of the JML estimator, as will be explained below. For the tetrad logit estimator, part (i) could be relaxed by assuming, instead, that $X_{ij}$ and $\sum_{k\in J}G_{ik}G_{jk}$ has a sufficient number of bounded moments and part (ii) of Assumption 3
is respect to network data can be difficult and expensive to collect; consequently, many analyses in the social sciences are based on incomplete graphs \citep{Banerjee2013}. One implication of part (ii) of Assumption 3 is that estimation and inference may be based upon only a subset of the full network \citep{Shalizi2013}.\\

 In the game with incomplete information, a sufficient condition for the existence of a single solution is given by the following assumption.

\begin{Asumm}(i) We have a sample i.i.d of $(G_{n_{\tau_{p}}}, X_{n_{\tau_{p}}})_{\tau_{p}\in T}$. Assume $T\to\infty$. (ii) $\bm{I}_{r,\tau_{p}}-\bm{\Phi}_{\tau_{p}}$ is nosingular, (iii) $\|\textbf{G}_{\tau_{p}}\|_{\infty}=1$, $\|\textbf{G}_{\tau_{p}}\|_{\max}=1$ for all $\tau_{p}\in T$ and  $\|\tilde{\bm{\Lambda}}_{\tau_{p}}\|_{1}<[\max_{k}\sup_{q}f_{k}(q)]^{-1}$ or $\|\tilde{\bm{\Lambda}}_{\tau_{p}}\|_{\infty}<[\max_{k}\sup_{q}f_{k}(q)]^{-1}$ and $\textbf{F}_{k,\tau_{p}}(\cdot)$ is a continuous distribution function supported compactly on $\mathbb{R}$ that is absolutely continuous with respect to the Lebesgue measure and with density function $f_{k}(\cdot)$.
\end{Asumm}
Assumption 4 suggests that, for the equilibrium to be unique, the social interaction effects cannot be too strong. So ensures that the first equation of the expression for the reduced form in \eqref{eq:reduced} is well defined. If $\tilde{\epsilon}_{im}^{*}$ follows the standard normal distribution, then
$\max_{k}\sup_{q} f_{k}(q)=1/\sqrt{2\pi}$. Furthermore, in some empirical studies of social networks, it
may be reasonable to have $\textbf{G}_{\tau_{p}}$ row-normalized \citep{Boucher2017,Lee2010}.
 In this case,
Assumption 4 holds if $\max_{k=1,\ldots,r}|\tilde{\lambda}_{lk}|<\sqrt{2\pi}$.
When $r=T=1$, Assumption 3 coincides with the sufficient condition for existence of
a unique rational expectation equilibrium for the single-activity social interaction model in \cite{Lee2010}.
\begin{prop}
If Assumption 4 holds, then the incomplete information network game with the utility \eqref{eq:Utility} has a unique pure strategy BNE with the equilibrium strategy profile.

\end{prop}
When the Assumption 4 is satisfied, the contraction $\textbf{g}(\bm{\psi}_{\tau_{p}})$  guarantee the consistency and completeness of the model \citep{Tamer2003}. However, some conditions omitted in the immediately mentioned works refer to transferable utility between tied players in any of its forms. In particular, to the efficiency and stability of such an equilibrium. The equilibrium concept we use is Pairwise Stability  \citep{Jackson1996}.

 The rule \eqref{eq:Gij} and utility \eqref{eq:Utility} guarantee us that the game is a game with transferable utility. We say that a network is pairwise stable if no pair of players want to create a new connection, and no player want to sever an existing link. Formally, we have the following
\begin{definition}\label{def_1}
	A network $G$  is pairwise stable (PS) under Transferable Utility if:\begin{itemize}
		\item For any $G_{ij}=1, \Delta\mathbb{E}_{w}(U_{ij}(G_{-ij},y^{*}_{ik})|R_{[j]})+\Delta\mathbb{E}_{w}(U_{ji}(G_{-ji},y^{*}_{jk}|R_{[j]})\geq 0$;
		\item For any $G_{ij}=0, \Delta\mathbb{E}_{w}(U_{ij}(G_{-ij},y^{*}_{ik})|R_{[j]})+\Delta\mathbb{E}_{w}(U_{ji}(G_{-ji},y^{*}_{jk}|R_{[j]}))< 0$;  
	\end{itemize}
\end{definition}
From remark \ref{rmk}, let $t^{i}_{ij}=\Delta\mathbb{E}_{w}(U_{ij}(G_{-ij},y^{*}_{ik})| R_{[j]})$ and $g(t)=\{(i,j)\in G:t^{i}_{ij}+t_{ji}^{j}\geq 0\}$. 
Since utility is interdependent, the pairwise stability condition reduces to an updated discrete choice symmetric model, that is,
\begin{equation}\label{eq:binamodel}
	G_{ij}=\textbf{1}\biggl\{t^{i}_{ij}+t^{j}_{ji}\geq 0\biggr\} \forall i\neq j
\end{equation}
Then, how the utility of player $i$ is updated by setting its connections and transfers, is given by:
\begin{equation*}
	\pi_{i}(G,a)(g(t))=U_{i}(G,\textbf{a})-\sum_{(i,j)\in g(t)}t^{i}_{ij}\geq 0
\end{equation*}
In particular, when these $n_{\tau_{p}}$ players form links, and a PS network
 $G_{\tau_{p}} = (G_{ij,\tau_{p}})_{i\neq j}$ emerge. The linking game with transfers is easily interpreted. Players simultaneously announce a transfer for each possible link that they might form. If the transfer is positive, it represents the offer that the player makes to form the link. If the transfer is negative, it represents the demand that a player requests to form the link. Note that the offer may exceed the demand, $t^{i}_{ij}+t^{j}_{ji}>0$. In that case, we hold both players to their promises \citep{Bloch2007}.\footnote{\cite{Bloch2007} also define version of the transfer game where players can also make indirect transfers, and also where they can make the transfers contingent on which network is formed. In the indirect transfer game, every player $i$ announces a vector of transfers $t_{i}\in\mathbb{R}^{n(n-1)/2}$. The entries in the vector $t_{i}$ are given by $t_{i,jk}$, denoting the transfer that player $i$ puts on the link $jk$. If $i\notin {jk}$, $t_{i,jk}\geq 0$. Player $i$ can make demands on the links that he or she is\textbf{} involved with, but can only make offers on the other links. Link ${jk}$ is formed if and only if $\sum_{i\in G}t_{i,jk}\geq 0$.}

  According to \cite{Jackson2002e}, for any utility function, there exists a PS network or a closed cycle.\footnote{A closed cycle is a collection of two or more distinct networks such that (i) for any two networks in the collection, there exists an improvement path from one to the other; and (ii) no improvement path starting from a network in the collection leads to an external network. Here, an enhancement path is a sequence of networks in which two consecutive networks differ in a link, and adding (or removing) the link in the successive network is beneficial to the individuals involved. See \cite{Jackson2002e} for rigorous definitions.}
\begin{prop}
	Suppose that the utility function is as in \eqref{eq:Utility}. Under TU, for any constants $\delta$ in $\mathbb{R}$, there is no closed cycle, so a PS network must exists.
\end{prop}
\cite{Sheng2020} showed that the existence of stable peer-to-peer networks is not guaranteed either. According to Lemma 1 of \cite{Jackson2001} for any utility function there is either a PS network or a closed
cycle. Intuitively a closed cycle represents a situation in which for the given utilities individuals never
reach a stable state and constantly switch between forming and severing links, which is unlikely to occur in real applications.\\
Most results in the network literature on the existence of PS networks do not allow for
heterogeneity among individuals and thus are unsuitable for our analysis. \cite{Jackson2001}, \cite{Bloch2006} \cite{Bloch2007} provided general conditions under which a PS network exists. The idea is applies their conditions and provide existence results for the utility function in \eqref{eq:Utility}.
 
\section*{\large Econometric Model and Identifications of Structural Parameters}

\subsection*{\small Econometric Model}

Consider a dataset containing $T$ networks. Our specification of the econometric model fllows closely from the equilibrium best response function of the theorical model. For  the $\tau_{p}$-th network, the best response functions is:

\begin{equation}\label{eq:best}
\textbf{y}_{m,\tau^{*}_{p}}=\sum_{l=1, l\neq m}^{r}\phi_{lm,\tau_{p}}\textbf{y}_{l,\tau^{*}_{p}}+\sum_{l=1}^{r}\sum_{\tau_{p}=1}^{T}\lambda_{lm,\tau_{p}}\textbf{G}_{\tau_{p}}\bm{\psi}_{l,\tau_{p}}+\overline{\textbf{w}}_{m,\tau_{p}}-\overline{\bm{\varepsilon}}_{m,\tau_{p}}
\end{equation}
Let $\overline{\textbf{w}}_{m,\tau_{p}}=\textbf{X}_{\tau_{p}}\bm{\alpha}_{m,\tau_{p}}+\bm{\iota}_{\tau_{p}}\bm{\gamma}_{m,\tau_{p}}$ where $\textbf{X}_{\tau_{p}}$ is a $n_{\tau_{p}}\times q$ matrix of observations on $q$ exogenous player characteristics, $\iota_{\tau_{p}}$ is an $n_{\tau_{p}}\times 1$ vector of ones. The, substitution of $\overline{\textbf{w}}_{k,\tau_{p}}$ into the best response functions gives the simultaneous equation netowrk model:
\begin{equation}\label{eq:subs}
	\textbf{y}_{m,\tau^{*}_{p}}=\sum_{l=1, l\neq m}^{r}\phi_{lm,\tau_{p}}\textbf{y}_{l,\tau^{*}_{p}}+\sum_{l=1}^{r}\sum_{\tau_{p}=1}^{T}\lambda_{lm,\tau_{p}}\textbf{G}_{\tau_{p}}\bm{\psi}_{l,\tau_{p}}+\textbf{X}_{\tau_{p}}\bm{\alpha}_{m,\tau_{p}}+\iota_{\tau_{p}}\bm{\gamma}_{m,\tau_{p}}-\bm{\overline{\epsilon}}_{m,\tau_{p}}
\end{equation}
in form more compact
\begin{equation}\label{eq:subs1}
	\textbf{y}^{*}_{m}=\sum_{l=1 ,l\neq m}^{r}\phi_{lm}\textbf{y}^{*}_{l}+\sum_{l=1}^{r}\lambda_{lm}\textbf{G}\bm{\psi}_{l}+\textbf{X}\bm{\alpha}_{m}+\textbf{L}\bm{\gamma}_{m}-\bm{\overline{\epsilon}}_{m}
\end{equation}
where \begin{align*}
	\textbf{y}^{*}_{m}=(\textbf{y}^{*}_{m,1},\ldots, \textbf{y}^{*}_{m,T})' \quad \textbf{G}=\diag\{\textbf{G}_{\tau_{p}}\}_{\tau_{p}=1}^{T}\quad
	\textbf{X}=(\textbf{X}'_{1},\ldots, \textbf{X}'_{T})'\quad \textbf{L}=\diag\{\iota_{\tau_{p}}\}_{\tau_{p}=1}^{T}\\
	\bm{\alpha}_{m}=(\bm{\alpha}_{m,1},\ldots, \bm{\alpha}_{m,T})\quad \bm{\gamma}_{m}=(\bm{\gamma}_{m,1},\ldots, \bm{\gamma}_{m,T})\quad
	\bm{\overline{\epsilon}}_{m}=(\bm{\overline{\epsilon}}_{m,1},\ldots,\bm{\overline{\epsilon}}_{m,T})
\end{align*}
 Now
\begin{equation}\label{eq:model_reduced}
	\textbf{Y}=\textbf{G}\bm{\Psi} \tilde{\bm{\Lambda}} + \textbf{X}\tilde{\bm{\alpha}}+\textbf{L}\tilde{\bm{\gamma}}-\bm{\tilde{\epsilon}}
\end{equation} 
with $\tilde{\bm{\alpha}}=\bm{\alpha}(\textbf{I}_{r}-\bm{\Theta})^{-1}$, $\bm{\tilde{\epsilon}}=\bm{\overline{\epsilon}}(\textbf{I}_{r}-\bm{\Theta})^{-1}$ and $\tilde{\bm{\gamma}}=\bm{\gamma }(\textbf{I}_{r}-\bm{\Theta})^{-1}$.\footnote{With a poyector $\textbf{P}=\diag\{\textbf{P}_{\tau_{p}}\}_{\tau_{p}=1}^{\overline{\tau}_{p}}$, where $\textbf{P}_{\tau_{p}}=\textbf{I}_{n_{\tau_{p}}}-\dfrac{1}{n_{\tau_{p}}}\iota_{n_{\tau_{p}}}\bm{\iota}_{n_{\tau_{p}}}$. This transformation in analogous to be the within transformation for fixed-effect panel data models. As $\textbf{P}\textbf{L}=0$, the within-transformed model is:
	\[
	\textbf{P}\textbf{y}_{m}=\sum_{l=1, l\neq m}^{r}\phi_{lm}\textbf{P}\textbf{y}_{l}+\sum_{l=1}^{r}\lambda_{lm}\textbf{P}\textbf{G}\bm{\psi}_{l}+\textbf{P}\textbf{X}\bm{\alpha}_{m}-\textbf{P}\bm{\overline{\epsilon}}_{m}
\] } We assume that $\vect(\bm{\tilde{\epsilon}})|\textbf{X}\sim \mathcal{N}(0,\bm{\Sigma}\otimes \textbf{I}_{n})$, where $\bm{\Sigma}=[\sigma_{lk}]$ is a $r\times r$ covariance matrix with $\sigma_{mm}=1$.

\subsection*{\small Identification of Structural parameters}

We consider the identification of the reduced form parameters $\bm{\Gamma}=[\tilde{\bm{\Lambda}}, \tilde{\bm{\alpha}}, \tilde{\bm{\gamma}}]'$. Let $\tilde{\lambda}_{lk}$ denote the $(l,k)$-th element of $\tilde{\bm{\Lambda}}$, $\tilde{\alpha}_{k}$ denote the $k$-th column of $\tilde{\bm{\alpha}}$ and $\tilde{\gamma}_{k}$ the $k$-th column of $\bm{\gamma}$.  Given the observed
adjacency matrix $\textbf{G}_{\tau^{*}_{p}}$ and exogenous covariates $\textbf{X}_{\tau^{*}_{p}}$, the parameters
$\bm{\Gamma}_{\tau_{p}}=[\tilde{\bm{\Lambda}}, \tilde{\bm{\alpha}}, \tilde{\bm{\gamma}}]_{\tau_{p}}'$ and the alternative parameters $\bm{\Gamma}_{\tau_{p}}'=[\tilde{\tilde{\bm{\Lambda}}}, \tilde{\tilde{\bm{\alpha}}}_{\tau_{p}}, \tilde{\tilde{\bm{\gamma}}}]_{\tau_{p}}'$ are observational equivalent if

\begin{equation}\label{eq:1}
	\begin{split}
	\mathbb{P}(a_{im,\tau_{p}}=1|\textbf{G}_{\tau^{*}_{p}},\textbf{X}_{\tau^{*}_{p}})&=\mathcal{N}(\sum_{l=1}^{r}\sum_{\tau_{p}=1}^{T}\tilde{\lambda}_{lm,\tau_{p}}\sum_{j=1}^{n}G_{ij,\tau_{p}}\psi_{jl,\tau_{p}}+\textbf{x}'_{i,\tau^{*}_{p}}\tilde{\bm{\alpha}}_{m,\tau_{p}}+\bm{\iota}_{\tau^{*_{p}}}\tilde{\bm{\gamma}}_{m,\tau_{p}})\\
	&=\mathcal{N}(\sum_{l=1}^{r}\sum_{\tau_{p}=1}^{T}\tilde{\tilde{\lambda}}_{lm,\tau_{p}}\sum_{j=1}^{n}G_{ij,\tau_{p}}\tilde{\psi}_{jl,\tau_{p}}+\textbf{x}'_{i,\tau^{*}_{p}}\tilde{\tilde{\bm{\alpha}}}_{m,\tau^{*}_{p}}+\bm{\iota}_{\tau^{*}_{p}}\tilde{\tilde{\bm{\gamma}}}_{m,\tau^{*}_{p}})
\end{split}
\end{equation}
for all $i=1,\ldots, n$ and $m=1,\ldots, r$, where, under Assumption 4, $\psi_{im,\tau_{p}}$, $\tilde{\psi}_{im,\tau_{p}}$ are uniquely determined by the fixed point mappings
\begin{equation}\label{eq:2}\psi_{im,\tau_{p}}=\mathcal{N}(\sum_{l=1}^{r}\sum_{\tau_{p}=1}^{T}\tilde{\lambda}_{lk,\tau_{p}}\sum_{j=1}^{n}G_{ij,\tau_{p}}\psi_{jl,\tau_{p}}+\textbf{x}'_{i,\tau^{*}_{p}}\tilde{\bm{\alpha}}_{m,\tau^{*}_{p}}+\bm{\iota}_{\tau^{*}_{p}}\tilde{\bm{\gamma}}_{m,\tau^{*}_{p}})
\end{equation}
\begin{equation}\label{eq:3}
	\tilde{\psi}_{im,\tau_{p}}=\mathcal{N}(\sum_{l=1}^{r}\sum_{\tau_{p}=1}^{\overline{\tau}_{p}}\tilde{\tilde{\lambda}}_{lm,\tau_{p}}\sum_{j=1}^{n}G_{ij,\tau_{p}}\tilde{\psi}_{jl,\tau_{p}}+\textbf{x}'_{i,\tau^{*}_{p}}\tilde{\tilde{\bm{\alpha}}}_{m,\tau^{*}_{p}}+\bm{\iota}_{\tau_{p}}\tilde{\tilde{\bm{\gamma}}}_{m,\tau^{*}_{p}})
\end{equation}
respectively. If \eqref{eq:1}, \eqref{eq:2} and \eqref{eq:3} hold, then 
\[\sum_{l=1}^{r}\sum_{\tau_{p}=1}^{\overline{\tau}_{p}}\Big(\tilde{\lambda}_{lm,\tau_{p}}-\tilde{\tilde{\lambda}}_{lm,\tau_{p}}\Big)\sum_{j=1}^{n}G_{ij,\tau_{p}}\psi_{jl,\tau_{p}}+\textbf{x}'_{i,\tau^{*}_{p}}\Big(\tilde{\bm{\alpha}}_{m,\tau^{*}_{p}}-\tilde{\tilde{\bm{\alpha}}}_{m,\tau^{*}_{p}}\Big)+\iota\Big(\tilde{\bm{\gamma}}_{m,\tau^{*}_{p}}-\iota\tilde{\tilde{\bm{\gamma}}}_{m,\tau^{*}_{p}}\Big)=0\]
More precisely

\begin{equation*}
	[\textbf{G}_{\tau_{p}}\bm{\Psi}_{\tau_{p}},\textbf{X}_{\tau_{p}}]\Big(\bm{\Gamma}_{\tau_{p}}-\bm{\Gamma}'_{\tau_{p}}\Big)=0
\end{equation*}
If each $[\textbf{G}_{\tau_{p}}\bm{\Psi}_{\tau_{p}},\textbf{X}_{\tau_{p}}]$ has full column rank, then the observational equivalence of $\bm{\Gamma}_{\tau_{p}}$ and $\tilde{\bm{\Gamma}}_{\tau_{p}}$ implies that are equals, i.e, the reduced form parameters is identifiable.

With the reduced form parameters identified, the structural parameters $\bm{\Theta}_{\tau_{p}}$, $\bm{\Lambda}_{\tau_{p}}$ and $\bm{\alpha}_{\tau_{p}}$ can be
identified from the equations $\tilde{\bm{\Lambda}}_{\tau_{p}}=\bm{\Lambda}_{\tau_{p}} (\textbf{I}_{r,\tau_{p}}-\bm{\Theta}_{\tau_{p}})^{-1}$ and $\tilde{\textbf{W}}_{\tau_{p}}=\textbf{W}_{\tau_{p}}(\textbf{I}_{r,\tau_{p}}-\bm{\Theta}_{\tau_{p}})^{-1}$. Suppose that, $k=1,\ldots,r$, there are $h_{k,\tau_{p}}$ restrictions on the $k$-th column, $\omega_{k,\tau_{p}}$, of $\bm{\Omega}_{\tau_{p}}=[\bm{\Theta}'_{\tau_{p}},-\bm{\Lambda}'_{\tau_{p}},-\bm{\alpha}'_{\tau_{p}}, \bm{\gamma}'_{\tau_{p}}]'$ of the form $\textbf{R}_{k,\tau_{p}}\omega_{k,\tau_{p}}=0$, where is a matrix of known constants. Then, the sufficient and necessary \textit{rank} condition for  the identification of the structural parameters $\bm{\Omega}$ from the
reduced form parameters is rank$(\textbf{R}_{k,\tau_{p}}\bm{\Omega}_{\tau_{p}})=r-1$ for $k=1,\ldots, r$, and the nedessary \textit{order} condition is $h_{k,\tau_{p}}\geq 1$ \citep{Rothenberg1971}.
\begin{Asumm}
	$[\textbf{G}_{\tau_{p}}\bm{\Psi}_{\tau_{p}}, \textbf{X}_{\tau_{p}}]$ has full column rank and rank$(\textbf{R}_{k,\tau_{p}}\omega_{k,\tau_{p}})=r-1$ for all $\tau_{p}=1,\ldots, T$.
\end{Asumm}
\begin{exam}
	Consider the model
	\begin{align*}
		\textbf{y}_{1}=&\phi_{21}\textbf{y}_{2}+\tilde{\lambda}_{21}\textbf{G}\psi_{2}+\textbf{X}\bm{\alpha}_{1}+\textbf{L}\bm{\gamma}_{1}-\bm{\epsilon}_{1}\\
		\textbf{y}_{2}=&\tilde{\lambda}_{12}\textbf{G}\psi_{1}+\textbf{X}\bm{\alpha}_{2}+\textbf{L}\bm{\gamma}_{2}-\bm{\epsilon}_{2}
	\end{align*}
	In this case, an player’s choice is influenced by the expectation on the peers’ choices in the same action. Furthermore, an player’s choice in action 1 is influenced by his/her own choice in action 2, but not the other way around. The exclusion restrictions $\tilde{\lambda}_{11}=\tilde{\lambda}_{22}=0$ can be written as $\textbf{R}_{2}\omega_{2}=0$ where 
	\[\textbf{R}_{2}=\begin{pmatrix}
		1&0&0&0&0&0_{1\times q}\\
		0&0&0&-1&0&0_{1\times q}
	\end{pmatrix}\]
	and $\omega_{2}=(\phi_{12},,\phi_{22},-\lambda_{12},-\lambda_{22},-\bm{\alpha}_{2},-\bm{\gamma}_{2})$. Similarly, the exclusion restriction $\tilde{\lambda}_{11}=0$ acn be write as $\textbf{R}_{1}\omega_{1}=0$ where 
	$\textbf{R}_{1}=(0,0,-1,0,0,0_{1\times q})$	
	and $\omega _{1}=(\phi_{11},\phi_{21},-\tilde{\lambda}_{11},\tilde{\lambda}_{21},-\bm{\alpha}_{1},-\bm{\alpha}_{2})$. Then
	
	\[\textbf{R}_{2}\bm{\Gamma}=\begin{pmatrix}
		1&\tilde{\lambda}_{21}\\
		0&0
	\end{pmatrix}\]
	has rank one and $\textbf{R}_{1}\bm=[0, -\tilde{\lambda}_{12}]$ wich also has rank one if $\tilde{\lambda}_{12}\neq 0$. If $\tilde{\lambda}_{12}= 0$, then the model becomes a classical linear simultaneous equation model, wich cannot be identified without imposing additional exclusion restrictions. If Assumption 6 holds, the structural parameters can be identified if the reduced form parameters are identifiable. The reduced form of the model is
	\begin{align*}
			\textbf{y}_{1}=&\lambda^{*}_{12}\textbf{G}\psi_{1}+\lambda^{*}_{21}\textbf{G}\psi_{2}+\textbf{X}\bm{\alpha}^{*}_{1}+\textbf{L}\bm{\gamma}^{*}_{1}-\bm{\epsilon}^{*}_{1}\\
		\textbf{y}_{2}=&\tilde{\lambda}^{*}_{12}\textbf{G}\psi_{1}+\textbf{X}\bm{\alpha}^{*}_{2}+\textbf{L}\bm{\gamma}^{*}_{2}-\bm{\epsilon}^{*}_{2}	
\end{align*}
	where \[\lambda^{*}_{12}=\phi_{21}\tilde{\lambda}_{12},\quad \lambda^{*}_{21}=\tilde{\lambda}_{21}\quad \bm{\alpha}^{*}_{1}=\bm{\alpha}_{1}+\bm{\alpha}_{2}\phi_{21}\quad \bm{\gamma}^{*}_{1}=\bm{\gamma}_{1}+\bm{\gamma}_{2}\phi_{21}\quad \tilde{\lambda}^{*}_{12}=\tilde{\lambda}_{12}\quad \bm{\gamma}^{*}_{2}=\bm{\gamma}_{2}\quad \bm{\gamma}^{*}_{2}=\bm{\gamma}_{2}   \]
	
\end{exam}
The Manski reflection problem arises from the coexistence of the within-action peer effect (also known as the endogenous peer effect in single-activity social interaction models) and the contextual effect
endogenous peer effect in single-activity models of social interaction) and the contextual effect. In Manski's linear-in-means
model, individuals are assumed to be affected by all members of their group and by no one outside the group, and therefore the simultaneity in the behavior of individuals in the same group introduces perfect collinearity between the within-activity peer effect and the contextual effect. Therefore, these two effects cannot be identified in the linear model at the means. However, in the vast majority of the work on social networks \citep{Acemoglu2012,Arieli2017,Golub2010,Goyal2009}, people are not affected equally by all members of the network. Instead, they are influenced by their (direct) connections. Therefore, the structure of social networks can be exploited to identify peer effects. This was originally recognized in \cite{Cohen2006} and systematically explored in \cite{Bramoulle2009}. Intuitively, if
individuals $i$ and $j$ are connected and $j$ and $k$ are connected, it does not necessarily follow that $i$ and $k$ are also connected. Because
the characteristics of the indirect connections of a player type are not collinear with its characteristics and the characteristics of its direct connections. Therefore, the characteristics of a player's indirect connections can be used as \textit{instruments} to identify the endogenous pair effect within the activity of the exogenous contextual effect. In particular, the expression \eqref{eq:model_reduced} identifies the exogenous contextual effect of the endogenous effect.
\subsection*{\small Bounds From Subnetworks and Game Types}

 For a given attribute profile $X_{\tau_{p}}$ and preference profile $\epsilon_{\tau_{p}}$, the model yields a collection of PS networks, denoted by $\mathcal{PS}(\Delta \mathbb{E}( U_{\tau_{p}}(X_{\tau_{p}},\epsilon_{\tau_{p}})))$, where $\Delta \mathbb{E}(U_{\tau_{p}}(X_{\tau_{p}},\epsilon_{\tau_{p}}))=\bigcap_{G_{-ij,\tau_{p}}\in \mathcal{G}_{-ij,\tau_{p}}, i\neq j}\{\Delta\mathbb{E}(U_{ij}(G_{-ij,\tau_{p}}, X_{ij,\tau_{p}},\epsilon_{ij,\tau_{p}}))\}$ represents the "marginal utility" profile and $\mathcal{G}_{-ij}$ represents all possible networks by removing the link of player $i$ with player $j$. To complete the model, suppose there is an equilibrium
 selection mechanism that selects a network from the collection of PS networks. Let $\sigma_{\tau_{p}}(\mathcal{PS}(\Delta\mathbb{E}(U_{\tau_{p}}(X_{\tau_{p}},\epsilon_{\tau_{p}})))$ be decision rule wich prescribes an subnetwork given her own private information from the PS collection. Then, conditional on $X_{\tau_{p}}$, the probability of observing the network $g_{\tau_{p}}$ is
 \begin{equation}\label{eq:int}
 	\mathbb{P}(G_{\tau_{p}}=g_{\tau_{p}}|X_{\tau_{p}})=\int \textbf{1}\Big\{\sigma_{\tau_{p}}(\mathcal{PS}(\Delta\mathbb{E}(U_{\tau_{p}}(X_{\tau_{p}},\epsilon_{\tau_{p}}))))=g_{\tau_{p}}\Big\}d \textbf{F}(\epsilon_{\tau_{p}})
 \end{equation}
The equation \eqref{eq:int} is similar in form but differs in content to \cite{Xiao2018i} and \cite{De2012} for discrete games with incomplete information.

Following \cite{Ciliberto2009}, the integral \eqref{eq:int} can be written as:
\begin{equation*}
	\mathbb{P}(G_{\tau_{p}}=g_{\tau_{p}}|X_{\tau_{p}})=\int_{g_{\tau_{p}}\in\mathcal{PS}(\Delta \mathbb{E}( U_{\tau_{p}}(X_{\tau_{p}},\epsilon_{\tau_{p}})))\&|\mathcal{PS}(\Delta \mathbb{E}( U_{\tau_{p}}(X_{\tau_{p}},\epsilon_{\tau_{p}}))|=1}d\textbf{F}(\epsilon_{\tau_{p}})
\end{equation*}
An immediate result is:
\begin{equation*}
	\mathbb{P}(G_{\tau_{p}}=g_{\tau_{p}}|X_{\tau_{p}})\leq \int_{g_{\tau_{p}}\in\mathcal{PS}(\Delta \mathbb{E}( U_{\tau_{p}}(X_{\tau_{p}},\epsilon_{\tau_{p}})))}d\textbf{F}(\epsilon_{\tau_{p}})
\end{equation*}
and \begin{equation*}
	\mathbb{P}(G_{\tau_{p}}=g_{\tau_{p}}|X_{\tau_{p}})\geq\int_{g_{\tau_{p}}\in\mathcal{PS}(\Delta \mathbb{E}( U_{\tau_{p}}(X_{\tau_{p}},\epsilon_{\tau_{p}})))\&|\mathcal{PS}(\Delta \mathbb{E}( U_{\tau_{p}}(X_{\tau_{p}},\epsilon_{\tau_{p}})))|=1}d\textbf{F}(\epsilon_{\tau_{p}})
\end{equation*}
For any $n_{\tau_{p}}\leq n, G=(G_{\tau_{p}},  G^{c}_{\tau_{p}})$. Thus the probability of observing a subnetwork $g_{\tau_{p}}$ is
\begin{equation*}
	\begin{split}
	\mathbb{P}(G_{\tau_{p}}=g_{\tau_{p}}|X_{\tau_{p}})=\int \sum_{g^{c}_{\tau_{p}}}\textbf{1}\Big\{\sigma_{\tau_{p}}( g^{c}_{\tau_{p}},\mathcal{PS}(\Delta\mathbb{E}(U_{n}(X_{n},\epsilon_{n}), X_{n},\epsilon_{n})=g_{\tau_{p}}\Big\}d \textbf{F}(\epsilon_{n})
	\end{split}
\end{equation*}
As in \cite{Sheng2020}, define
\[\mathcal{PS}_{g_{\tau_{p}}}(\Delta U_{n}(X_{n},\epsilon_{n}))=\big\{g_{\tau_{p}}:\exists g^{c}_{\tau_{p}}, (g_{\tau_{p}},g^{c}_{\tau_{p}})\in \mathcal{PS}(\Delta U_{n}(X_{n},\epsilon_{n}))\big\}\]
to be the set of networks in $g_{\tau_{p}}$ that can be part of a PS network in the set $\mathcal{PS}(\Delta U_{n}(X_{n},\epsilon_{n}))$.
\begin{lemm}
	For any subnetwork $g_{\tau_{p}}$ with $n_{\tau_{p}}\leq n$, the subnetwork choice probability $\mathbb{P}(G_{\tau_{p}}=g_{\tau_{p}}|X_{\tau_{p}})$ is bounded by
	\begin{equation}\label{eq:bond}
		LB(g_{\tau_{p}},X_{\tau_{p}})\leq \mathbb{P}(G_{\tau_{p}}=g_{\tau_{p}}|X_{n})\leq UB(g_{\tau_{p}},X_{\tau_{p}})
	\end{equation}
with
\[
	UB(g_{\tau_{p}},X_{n})=\int_{g_{\tau_{p}}\in\mathcal{PS}_{\tau_{p}}(\Delta \mathbb{E}(U_{n}(X_{n},\epsilon_{n})))}d\textbf{F}(\epsilon_{n})\]
	\[
	LB(g_{\tau_{p}},X_{n})=\int_{g_{\tau_{p}}\in\mathcal{PS}_{\tau_{p}}(\Delta U_{n}(X_{n},\epsilon_{n}))\&|\mathcal{PS}(\Delta \mathbb{E}( U_{n}(X_{n},\epsilon_{n})))|=1}d\textbf{F}(\epsilon_{n})\]
\end{lemm}
When we observese at a subnet $g_{\tau_{p}}$, it must be part of a PS network for some complement
$g^{c}_{\tau_{p}}$ because otherwise definition \eqref{def_1} would not be satisfied, in turn, such PS network $(g_{\tau_{p}}, g^{c}_{\tau_{p}})$ must be selected by the equilibrium selection mechanism. Without information about the equilibrium selection, the probability
of observing subnetwork $g_{\tau_{p}}$ is bounded above by the probability that such subnetwork can be part of a PS network for some complement $g^{c}_{\tau_{p}}$, which gives
the upper bound in \eqref{eq:bond}, and bounded below by the probability that only the subnetwork $g_{\tau_{p}}$ can be part of a PS network for some complement $g^{c}_{\tau_{p}}$, which gives the smallest
bounded in \eqref{eq:bond}. 
	
\section*{\large Estimations}
{\textbf{E.1 Joint Maximum Likelihood Estimation} }

Let $\textbf{A}_{n}$ denote an $n \times1$ vector of degree heterogeneity values and $\textbf{A}_{n_{0}}$ the corresponding vector of true values. For what follows it is also convenient to define the notation
\[p_{ij}(\theta,\textbf{A})=\dfrac{\exp(\big(\sum_{k\in J}G_{ik}G_{jk}\big)\delta+X_{ij}'\beta+\iota'_{ij}\textbf{A})}{1+\exp\big(\big(\sum_{k\in J}G_{ik}G_{jk}\big)\delta+X_{ij}'\beta+\iota'_{ij}\textbf{A}\big)}\]
Let be $\theta=(\beta,\delta)$. The joint maximun likelihood choose $\title{\theta_{JML}}$ and $\tilde{\textbf{A}}_{n}$ simultaneously in order to maximize the log-likelihood fuction
\begin{equation}\label{eq:Log}
	L(\theta,\textbf{A}_{n})=\sum_{i< j}\biggl\{G_{ij}\ln p_{ij}(\theta,\textbf{A})+(1-G_{ij})\ln [1-p_{ij}(\theta,\textbf{A})]\biggr\}
\end{equation}
For this purpose,
it is convenient to note that $\tilde{\theta}_{JML}$ also maximizes the concentrated likelihood
\begin{equation}\label{eq:Log_concentred}
	L(\theta,\tilde{\textbf{A}}(\theta))=\sum_{i< j}\biggl\{G_{ij}\ln p_{ij}(\theta,\tilde{\textbf{A}}_{n}(\theta))+(1-G_{ij})\ln [1-p_{ij}(\theta,\tilde{\textbf{A}}_{n}(\theta))]\biggr\}
\end{equation}
where $\tilde{\textbf{A}}_{n}(\theta)=\arg\max_{A\subset \mathbb{R}^{n}}L(\theta,\textbf{A})$.

By adapting Theorem 1.5 of \cite{Chatterjee2011}, we show that $\tilde{\textbf{A}}_{n}$, when exists, is the unique solution to the fixed point problem
\begin{equation}
	\tilde{\textbf{A}}_{n}(\theta)=\phi(\tilde{\textbf{A}}_{n}(\theta)),
\end{equation}
where
\[\phi(\textbf{A})=\begin{pmatrix}
	\ln G_{1+}-\ln r_{1}(\theta,\textbf{A},\textbf{G}_{1})\\
	\ln G_{2+}-\ln r_{2}(\theta,\textbf{A},\textbf{G}_{2})\\
	\vdots\\
	\ln G_{n+}-\ln r_{n}(\theta,\textbf{A},\textbf{G}_{n})
\end{pmatrix}\]
with $\textbf{G}_{i}=(G_{i1},\ldots, G_{i(i-1)},G_{i(i+1)},\ldots,G_{in})'$ and
\[r_{i}(\theta,\textbf{A}(\theta),\textbf{G}_{i})=\sum_{i\neq j}\dfrac{\exp\big(\big(\sum_{k\in J}G_{ik}G_{jk}\big)\delta+X_{ij}'\beta\big)}{\exp(-A_{j}(\theta))+\exp\big(\big(\sum_{k\in J}G_{ik}G_{jk}\big)\delta+X_{ij}'\beta+A_{i}(\theta)\big)}\]
That $\tilde{\textbf{A}}_{n}(\theta)=\phi(\tilde{\textbf{A}}_{n}(\theta))$ can be directly verified by rearranging the sample score of \eqref{eq:Log}. That iteration using 	$\tilde{\textbf{A}}_{n}(\theta)=\phi(\tilde{\textbf{A}}_{n}(\theta))$ converges to $\tilde{\textbf{A}}_{n}(\theta)=\arg\max_{A\subset \mathbb{R}^{n}}L(\theta,\textbf{A})$ when the solution exists.

The fixed point representation of $\tilde{\textbf{A}}_{n}(\theta)$ shows that, while the incidental parameters $\{A_{i}\}_{i=1}^{n}$ are player-specific, their concentrated MLE are jointly determined using all $\binom{n}{2}$ link observations. To see this, observe that if we perturb $\tilde{A}_{i}$ , then all values of $\tilde{A}_{i}$ for
$i\neq j$ will change. This differs from joint fixed effects estimation in a nonlinear panel data model without time effects. In those models, conditional on the common parameter, the value of $\tilde{A}_{i}(\theta)$ is a function of only the $T$ observations specific to unit $i$ \citep{Arellano2007,Graham2017}.
The joint determination of the components
of $\tilde{\textbf{A}}_{n}(\theta)$ is a direct consequence of the multi-player nature of the network formation
problem and complicates the asymptotic analysis of $\tilde{\theta}_{JML}$.

With Assumption 2, we observe that
\begin{equation}\label{eq:p_ij}
	p_{ij}(\theta,A_{i},A_{j})\in (\xi,1-\xi)
\end{equation}
for some $\xi\in(0,1)$ and for all $(A_{i},A_{j})\in \mathbb{I}^{2}$ and $\theta=(\beta,\delta)\in \mathbb{A}\times \mathbb{B}$. Condition \eqref{eq:p_ij} implies that, in large networks, the number of observed links per agent will be proportional to the number of sampled agents. Put differently, it implies a dense sequence of graphs \citep{Graham2017}. Second par od the assumption is an identification condition. It will generally hold if there is sufficient variance in each column of $\textbf{G}_{i}=(G_{i1},\ldots,G_{i(i-1)},G_{i(i+1)},\ldots, G_{in})'$.
\begin{prop}\label{pro3}
	Under Assumptions 1, 2 and 3 \[\tilde{\theta}_{JML}\xrightarrow{p} \theta_{0}\]
\end{prop}

{\textbf{E.2 Transitive structure}}

Three players $i, j$ and $k$ are in a transitive relationship if, possibly upon reshuffling the labels within the triad, the network contains the links $ij$, $ik$ and $jk$, The subnetwork $\vartheta:=\{ij, ik, jk\}$ is called a transitive triangle and its is observed if $\vartheta\subset \{ij\in G:G_{ij}=1\}$. The set of all transitive triangles on the complete network is given by $G(n)$.\footnote{The terms transitive triangle and cyclic triangle are adapted from the notion of transitive
	and cyclic triads in \cite{Holland1976}.} For every transitive triangle $\vartheta$ take $\vartheta=(\vartheta_{1},\vartheta_{2},\vartheta_{3})$. Let $G_{\vartheta}=\prod_{e\in \vartheta}G_{\vartheta_{1}}G_{\vartheta_{2}}G_{\vartheta_{3}}$ denote the binary indicator that takes the value one if $\vartheta$ is observed and the value zero otherwise. We can construct measures of network transitivity by counting the number of transitive triangles in the network:
\[S_{n}=\sum_{\vartheta\in G(n)}G_{\vartheta}.\]
Given vector of player characteristics $(X'_{i},A_{i})$ our best prediction of the observed number of transitive triangles is given by $\mathbb{E}(S_{n})$. The discrepancy between the observed and the predicted level of transitivity can be summarized by a measure of excess transitivity defined as
\begin{equation}\label{eq:oracle}
	E_{n}=\dfrac{S_{n}-\mathbb{E}(S_{n})}{n^{3}}
\end{equation}
where the denominator normalizes by the number of transitive triangles in the complete network, $|G(n)|=n^{3}$.\footnote{This measure of excess transitivity translates a concept for undirected networks discussed
	in \cite{Karlberg1997} to directed networks and recent studied by \cite{Dzemski2019} to indirected netwotks. An alternative is to standardize by the number
	of open triangles, yielding the clustering coefficient \cite{Jackson2007}.}
Positive values of this statistic indicate that we observe \textit{more} transitive
relationships than expected, negative values of the statistic indicate that we observe \textit{fewer} transitive relationships than expected. Under an asymptotic sequence of reference distributions that takes the number of players.

Under the dyadic linking model, the conditional probability of observing a transitive triangle $\vartheta \in G(n)$ is given by $\mathbb{E}[G_{\vartheta}]=\prod_{e\in\vartheta}p_{e}(\theta_{0},\textbf{A}_{0})$. The parameter $\theta_{0}$ are unknown and it is
not feasible to compute $\mathbb{E}[S_{n}]=\sum_{\vartheta\in G(n)}\mathbb{E}[G_{\vartheta}]$ in $E_{n}$.  A feasible test statistic is given by
\[\hat{E}_{n}=\dfrac{S_{n}-\widehat{\mathbb{E}[S_{n}]}}{n^{3}}\]
where we replaced $\mathbb{E}[S_{n}]$ by the na\.{i}ve plug-in estimator
\[\widehat{\mathbb{E}[S_{n}]}=\sum_{\vartheta\in G(n)}\prod_{e\in\vartheta}p_{e}(\hat{\theta},\textbf{A}).\]
A theoretical analysis of $\hat{E}_{n}$ can be based on the decomposition
\begin{equation}
	n\hat{E}_{n}=nE_{n}-n^{-2}\sum_{\vartheta\in G(n)}\Bigg(\prod_{e\in \vartheta}p_{e}(\hat{\theta},\textbf{A})-\prod_{e\in \vartheta}p_{e}(\theta_{0},\textbf{A}_{0})\Bigg)
\end{equation}

Both terms on the right-hand side are of the same stochastic order and contribute to the asymptotic distribution. The first term is the appropriately scaled oracle statistic. Under the dyadic linking model it is centered at zero. The second term represents the effect estimating linking probabilities. Because of the incidental parameter problem, this term is not centered at zero. Consequently, the sign of $\hat{E}_{n}$ cannot be interpreted in the same way as the sign of $E_{n}$. In particular, values of $\hat{E}_{n}$ that are close to zero do not indicate that the
observed level of transitivity is consistent with the true dyadic linking model.
In preparation for a formal analysis of $\hat{E}_{n}$, let
\[\bm{\chi}_{ij,n}=\dfrac{1}{nH_{ij}}\sum_{\vartheta\in G(n), ij\in \vartheta}\mathbb{E}[G_{\vartheta}|G_{ij}=1]\]
where $H_{ij}=\partial_{z}p_{ij}/p_{1,ij}$.\footnote{Write $p_{1,ij}=p_{ij}(1-p_{ij})$ for the conditional variance of $G_{ij}$, and for functions $g\mapsto p(g)$, write $p_{ij}=p(g_{ij})$, for example $g_{1}=g_{ij}=\big(\sum_{k\in J}G_{ij}G_{ik}G_{jk}\big)\delta+X_{ij}'\beta+A_{ij}$, $g_{2}=g_{ji}=\big(\sum_{k\in J}G_{ji}G_{jk}G_{ik}\big)\delta+X_{ji}'\beta+A_{ji}$ and $\partial_{g^{w}}p_{ij}=\partial_{g^{w}}p(g)|g=g_{ij}$ for $w\in \mathbb{N}$.} The sum in the definition of $\bm{\chi}_{ij,n}$ counts the expected number of observed triangles containing the link $ij$ conditional on observing $ij$. Let $\bm{\chi}_{n}=\{\bm{\chi}_{ij,n}\}_{ij\in G}$. Define the projected vector $\tilde{\bm{\chi}}_{n}=\bm{\chi}_{n}-\mathcal{P}\bm{\chi}_{n}$ where $\mathcal{P}$ denote the projection operator that orthogonally projects vectors, $\rho_{ij}=H_{ij}(\partial_{z}p_{ij})$ and $X_{l}$ denote the residual of the projection of $l$th  component of the link-specific covariate, i.e, $X_{l}=(X_{ij,l})_{ij\in G}$, $\tilde{X}_{l}=X_{l}-\mathcal{P}X_{l}$.\footnote{These quantities are linked to the score and the Hessian of the unconstrained maximum likelihood problem. In particular, writing $l_{ij}=G_{ij}\ln(p_{ij})+(1-G_{ij})\ln(1-p_{ij}) $
	for the likelihood contribution of link $ij$, we have $\partial_{z}l_{ij}=H_{ij}(G_{ij}-p_{ij})$ and $\mathbb{E}[-\partial_{z^{2}}l_{ij}]=\rho_{ij}$.} Let $\tilde{X}_{ij}$ denote te the column
vector $(\tilde{X}_{ij,1},\ldots, \tilde{X}_{ij,\dim(\beta)})'$ and $l_{ij}=G_{ij}\ln(p_{ij})+(1-G_{ij})\ln(1-p_{ij})$ so that we can write $L(\theta,\textbf{A})=\frac{1}{n}\sum_{i,j=1,i\neq j}^{n}l_{ij}(\theta,\textbf{A})$. Closely to \cite{Dzemski2019}, for $ij\in G$ define
\[s_{ij}(\theta,\varrho)=s(\big(\sum_{k\in J}G_{ik}G_{jk}\big)\delta+X'_{ij}\beta+A_{i}+A_{j},\big(\sum_{k\in J}G_{jk}G_{ik}\big)\delta+X'_{ji}\beta+A_{j}+A_{i},\varrho)\]
denote the distribution function of a bivariate normal random variable
with marginal variances equal to one and covariance $\varrho$. 
This function can be used to compute the conditional probability of observing a reciprocated link.
\begin{rmk}
	The shocks $(\nu_{ij},\nu_{ji})$  are drawn independently across dyads ij  from a bivariate normal
	distribution with covariance $\varrho_{0}$  and marginal variances equal to one. If $\varrho_{0}$ 
	is positive then players will tend to reciprocate link \citep{Dzemski2019,Hoff2005}.  Let $\overline{\mathbb{E}}$ denote the conditional expectation operator that integrates out the randomness in $(\nu_{ij})_{ij\in G}$. Then,
	\begin{align*}
		\overline{\mathbb{E}}[G_{ij}G_{ji}]=&\mathbb{P}\Bigg(\nu_{ij}\leq \big(\sum_{k\in J}G_{ik}G_{jk}\big)\delta+X'_{ij}\beta+A_{i}+A_{j},\nu_{ji}\leq\big( \sum_{k\in J}G_{jk}G_{ik}\big)\delta+X'_{ji}\beta+A_{j}+A_{i}|X_{i},X_{j},\varrho)\Bigg)\\
		=&s_{ij}(\theta_{0},\textbf{A}_{0},\varrho_{0})
		\end{align*}
	for $\tilde{c}\in (0,1/2)$ and 
	\[m_{ij}(\theta,\textbf{A},\varrho)=G_{ij}G_{ji}\ln(s_{ij}(\theta,\textbf{A},\varrho))+(1-G_{ij}G_{ji})\ln(1-s_{ij}(1-s_{ij}(\theta,\textbf{A},\varrho)))\]
	the estimator $\hat{\varrho}$ solves the maximization problem
	\begin{equation}
		\hat{\varrho}=\arg\max_{\varrho\in[-1+\tilde{c},1-\tilde{c}]}\dfrac{1}{n}\sum_{i=1}^{n}\sum_{j=1}^{n}m_{ij}(\hat{\theta},\hat{\textbf{A}},\varrho).
	\end{equation}
\end{rmk}

Let $\tilde{\varrho}_{ij}=(s_{ij}(\theta_{0},\textbf{A}_{0},\varrho_{0})-p_{ij}p_{ji})/\sqrt{p_{1,ij}p_{1,ji}}$ the conditional correlation between $G_{ji}$ and $G_{ji}$ and 
\[\text{corr}_{i}=\dfrac{\sum_{j=1,j\neq i}^{n}\tilde{\varrho}_{ij}\sqrt{\rho_{ij}\rho_{ji}}}{\Big(\sum_{j=1,j\neq i}^{n}\rho_{ij}\Big)^{1/2}\Bigg(\sum_{j=1,j\neq i}^{n}\rho_{ji}\Bigg)^{1/2}}\]
the measures the correlation of all $\partial_{g}l_{ij}$ in the neighborhood of player $i$. The following result establishes convergence of $\hat{E}_{n}$ to a normal limit and gives expressions for its asymptotic bias and variance.

\begin{lemm}(Behavior of stochastic part of $\hat{\varrho}$)
	Under Assumption 2 and 3. Let
	\[\mathcal{M}(\theta,\textbf{A},\varrho)=\dfrac{1}{n}\sum_{i=1}^{n}\sum_{j=1,j\neq i}^{n}m_{ij}(\theta,\textbf{A},\varrho),\quad J_{ij}=\partial_{\varrho}s_{ij}/s_{1,ij}, H(\theta,\textbf{A})=-\partial_{\textbf{A}\textbf{A}'}L(\theta,\textbf{A})\]
	Then
	\begin{equation}
		\partial_{\varrho^{2}}\mathcal{M}=-\dfrac{1}{n}\sum_{i=1}^{n}\sum_{j=1,i<j}^{n}J_{ij}(\partial_{\varrho}s_{ij})+O_{p}(1)
	\end{equation}
\begin{multline}
		\partial_{\varrho \delta}\mathcal{M}+\partial_{\varrho \beta}\mathcal{M}+(\partial_{\varrho A}\mathcal{M}\overline{H}^{-1}\partial_{\varrho \beta}\overline{L})=-\dfrac{1}{n}\sum_{i=1}^{n}\sum_{j=1,j\neq i}^{n}J_{ij}\Big(\sum_{k\in J}G_{ijk}(\partial_{g_{1}}s_{ij}+\partial_{g_{2}}s_{ij})+\\
		+X'_{ij}(\partial_{g_{1}}s_{ij})-s_{ij}\nabla'_{ij}\Big)+O_{p}(1)
\end{multline}
	where $G_{ijk}=G_{ij}G_{ik}G_{jk}$.
\end{lemm}

\begin{lemm}
	Under Assumption 2 and 3\[s_{n}(\hat{\theta},\hat{\textbf{A}})-s_{n}(\theta_{0},\textbf{A}_{0})=\Bigg[(\partial_{\beta'}s_{n})+(\partial_{\delta'}s_{n})+(\partial_{\textbf{A}}s_{n}H^{-1}(\partial_{\textbf{A}\beta\delta})L)\Bigg](\hat{\theta}-\theta_{0})+\partial_{\textbf{A}}s_{n}H^{-1}S+B_{1,n}+o(1)\]
	where $S=\partial_{\textbf{A}}L(\theta,\textbf{A})$ and
	\[B_{1,n}=\dfrac{1}{2}(\partial_{\textbf{A}'}s_{n})\sum_{h=1}^{\dim(A)}\partial_{\textbf{A}\textbf{A}'\textbf{A}}LH^{-1}S+\dfrac{1}{2}(H^{-1}S)'[\partial_{\textbf{A}\textbf{A}'}s_{n}]H^{-1}S\]
	\end{lemm}
\begin{prop}
	Let \[V_{n}=\dfrac{1}{n^{2}}\sum_{i=1}^{n}\sum_{j=1,j\neq i}^{n}\bm{\chi}_{ij,n}\rho_{ij}\tilde{X}_{ij},\quad \overline{W}_{1,n}=\dfrac{1}{n^{2}}\sum_{i=1}^{n}\sum_{j=1, j\neq i}^{n}\rho_{ij}\tilde{X}_{ij}\tilde{X}'_{ij}\]
	such that for $\lambda(M)=\min\{\lambda:M\lambda= \lambda x\}$ hold $\lim\inf _{n\to \infty} \lambda(W_{1,n})>0$ and $v_{ij,n}=V'_{n}W_{1,n}\tilde{X}_{ij}$. Then
	\begin{equation}
		\dfrac{n\hat{E}_{n}+B^{\hat{E}}_{n}+V'_{n}W^{-1}_{1,n}B^{\beta}_{n}+B^{\hat{E}}_{n}+V'_{n}W^{-1}_{1,n}B^{\delta}_{n}}{\sqrt{v^{\hat{E}}_{n}}}\xrightarrow{d}\mathcal{N}(0,1)+o(1),
	\end{equation}
where
\[B^{\delta}_{n}=\Bigg[\dfrac{1}{2\sqrt{n}}\sum_{i=1}^{n}\dfrac{\frac{1}{n-1}\sum_{j=1,j\neq i}^{n}\rho_{ij}\tilde{X}_{ij}\tilde{X}'_{ij}}{\frac{1}{n-1}\sum_{j=1,j\neq i}^{n}\rho_{ij}}\Bigg]\delta_{0}+\Bigg[\frac{1}{2\sqrt{n}}\sum_{j=1}^{n}\frac{\frac{1}{n-1}\sum_{i=1,i\neq j}^{n}\rho_{ij}\tilde{X}_{ij}\tilde{X}'_{ij}}{\frac{1}{n-1}\sum_{i=1,i\neq j}^{n}\rho_{ij}}\Bigg]\delta_{0},\]

\[B^{\beta}_{n}=\Bigg[\dfrac{1}{2\sqrt{n}}\sum_{i=1}^{n}\dfrac{\frac{1}{n-1}\sum_{j=1,j\neq i}^{n}\rho_{ij}\tilde{X}_{ij}\tilde{X}'_{ij}}{\frac{1}{n-1}\sum_{j=1,j\neq i}^{n}\rho_{ij}}\Bigg]\beta_{0}+\Bigg[\frac{1}{2\sqrt{n}}\sum_{j=1}^{n}\frac{\frac{1}{n-1}\sum_{i=1,i\neq j}^{n}\rho_{ij}\tilde{X}_{ij}\tilde{X}'_{ij}}{\frac{1}{n-1}\sum_{i=1,i\neq j}^{n}\rho_{ij}}\Bigg]\beta_{0},\]
 
\begin{multline*}
B^{\hat{E}}_{n}=\dfrac{1}{2\sqrt{n}}\sum_{i=1}^{n}\dfrac{\frac{1}{n-1}\sum_{j=1,j\neq i}^{n}H_{ij}(\partial_{z^{2}}p_{ij})\hat{\bm{\chi}}_{ij,n}}{\frac{1}{n-1}\sum_{j=1,j\neq i}^{n}\rho_{ij}}+\dfrac{1}{2\sqrt{n}}\sum_{i=1}^{n}\dfrac{\frac{1}{n-2}\sum_{k=1,k\neq i,j}^{n}(\partial_{z}p_{ij})(\partial_{z}p_{ik})[p_{jk}+p_{kj}]}{\frac{1}{n-1}\sum_{j=1,j\neq i}^{n}\rho_{ij}}+\\
+\dfrac{1}{2\sqrt{n}}\sum_{j=1}^{n}\dfrac{\frac{1}{n-1}\sum_{i=1,i\neq j}^{n}H_{ij}(\partial_{z^{2}}p_{ij})\hat{\bm{\chi}}_{ij,n}}{\frac{1}{n-1}\sum_{j=1,j\neq i}^{n}\rho_{ij}}+\dfrac{1}{2\sqrt{n}}\sum_{j=1}^{n}\dfrac{\frac{1}{n-2}\sum_{k=1,k\neq i,j}^{n}(\partial_{z}p_{ij})(\partial_{z}p_{kj})[p_{ik}+p_{ki}]}{\frac{1}{n-1}\sum_{i=1,i\neq j}^{n}\rho_{ij}}+\\
+\dfrac{1}{\sqrt{n}}\sum_{i=1}^{n}\dfrac{\frac{\text{corr}_{i}}{n-2}\sum_{j=1,j\neq i}^{n}\sum_{k=1,k\neq i,j}^{n}(\partial_{z}p_{ij})(\partial_{z}p_{ij})p_{kj}}{\Big(\frac{1}{n-1}\sum_{j=1,j\neq i}^{n}\rho_{ij}\Big)^{1/2}\Big(\frac{1}{n-1}\sum_{j=1,j\neq i}^{n}\rho_{ij}\Big)^{1/2}}
\end{multline*}
and
\[v^{\hat{E}}_{n}=\dfrac{1}{n^{2}}\sum_{i=1}^{n}\sum_{j=1,j\neq i}^{n}\Biggl[(\tilde{\bm{\chi}}_{ij,n}-v_{ij,n})^{2}\rho_{ij}+(\tilde{\bm{\chi}}_{ij,n}-v_{ij,n})(\tilde{\bm{\chi}}_{ji,n}-v_{ij,n})(\tilde{\varrho}_{ij}\sqrt{\rho_{ij}\rho_{ji}})\Biggr]\]

\end{prop}
If linking probabilities are sufficiently small $p_{ij}\leq 1/2$. If the link surplus does not contain a homophily component, then $\hat{E}_{n}$ will be centered at a negative
value if the dyadic linking model is the true model. In more general specifications, the sign
of the bias depends on the numerical values of the structural parameters and can be positive
or negative.
 Intuitively, $\hat{E}_{n}$ compares the observed transitivity against the transitivity predicted by the dyadic model that provides the best fit. Therefore, my test looks only at the variation in transitivity that cannot be explained by degree distributions that are spanned by the
sender and receiver effects.

\subsection*{\small E.3 Econometric model}

The estimation of model \eqref{eq:model_reduced} follows the above identification strategy. First, we estimate the
reduced form parameters $\bm{\Gamma}^{*}=[\tilde{\Lambda}^{*},\bm{\tilde{\alpha}}^{*},\tilde{\bm{\gamma}}^{*}]$ using the NPL algorithm. The NPL algorithm was proposed by \cite{Aguirregabiria2007} for the estimation of dynamic discrete-choice games,
and has recently been adopted by \cite{Liu2019}  for the estimation of large network games.\footnote{ \cite{Aguirregabiria2007} studies the estimation of dynamic discrete games of incomplete information. Two main econometric issues appear in the estimation of these models: the indeterminacy problem associated with the existence of multiple equilibria and the computational burden in the solution of the game. Also, propose a class of pseudo maximum likelihood (PML) estimators that deals with these problems, and study the asymptotic and finite sample properties of several estimators in this class. First focus on
	two-step PML estimators, which, although they are attractive for their computational
	simplicity, have some important limitations: they are seriously biased in small samples;
	they require consistent nonparametric estimators of players’ choice probabilities in the
	first step, which are not always available; and they are asymptotically inefficient. Second, we show that a recursive extension of the two-step PML, which we call nested
	pseudo likelihood (NPL), addresses those drawbacks at a relatively small additional
	computational cost. The NPL estimator is particularly useful in applications where consistent nonparametric estimates of choice probabilities either are not available or are
	very imprecise, e.g., models with permanent unobserved heterogeneity.
	\cite{Liu2019} discuss the identification of the econometric model and propose a two stage estimation procedure, where the reduced form parameters are estimated by the nested
	pseudo likelihood (NPL) algorithm \citep{Aguirregabiria2007} in the first stage and the
	structural parameters are recovered from the estimated reduced form parameters by the
	Amemiya generalized least squares (AGLS) estimator \citep{Amemiya1978} in the second stage.}
The NPL algorithm for the reduced form \eqref{eq:reduced} starts from an arbitrary initial value $\bm{\Psi}^{(0)}\in[0,1]^{nr}$ and  takes the following iterative steps:

\underline{Step 1}

Choose $\tau^{*}_{p}\geq 1$ and given $\bm{\Psi}_{\tau^{*}_{p}}^{(j-1)}$ obtain
\begin{multline*}
	\hat{\bm{\zeta}}^{*(j)}_{k,\tau^{*}_{p}}=(\hat{\lambda}_{1k,\tau^{*}_{p}}^{*(j)},\ldots, \hat{\lambda}_{rk,\tau^{*}_{p}}^{*(j)}, \hat{\bm{\alpha}}_{k,\tau^{*}_{p}}^{*(j)'},\hat{\bm{\gamma}}_{k,\tau^{*}_{p}}^{*(j)'})'=\arg \max \ln L(\bm{\zeta}_{k,\tau^{*}_{p}};\bm{\Psi}_{\tau^{*}_{p}}^{(j-1)})=\\
	=\sum_{i=1}^{n}a_{ik}\ln \mathcal{N}\Big(\sum_{l=1}^{r}\sum_{\tau_{p}=1}^{\overline{\tau_{p}}}\lambda_{lk,\tau_{p}}^{*}\sum_{j=1}^{n}G_{ij,\tau_{p}}\psi_{lk,\tau_{p}}^{(j-1)}+\textbf{x}'_{i,\tau_{p}}\bm{\alpha}^{*}_{k,\tau_{p}}+\bm{\iota}_{k,\tau_{p}}\bm{\gamma}_{k,\tau_{p}}^{*}\Big)+\\
	+\sum_{i=1}^{n}(1-a_{ik})\ln \Big[1-\mathcal{N}\Big(\sum_{l=1}^{r}\sum_{\tau_{p}=1}^{\overline{\tau_{p}}}\lambda^{*}_{lk,\tau_{p}}\sum_{j=1}^{n}G_{ij,\tau_{p}}\psi_{jl,\tau_{p}}^{(j-1)}+\textbf{x}'_{i,\tau_{p}}\bm{\alpha}^{*}_{k,\tau_{p}}+\bm{\iota}_{k,\tau_{p}}\bm{\gamma}^{*}_{k,\tau_{p}}\Big)\Big]
\end{multline*}
for $k=1,\ldots, r$.

\underline{Step 2}

Given $\hat{\bm{\Xi}}^{*(j)}_{\tau^{*}_{p}}=[\hat{\bm{\zeta}}^{*(j)}_{1,\tau\tau^{*}_{p}},\ldots, \hat{\bm{\zeta}}^{*(j)}_{r,\tau\tau^{*}_{p}}]$, obtain  
\[\bm{\Psi}^{(j)}_{\tau\tau^{*}_{p}}=\textbf{g}(\bm{\Psi}^{(j-1)}_{\tau^{*}_{p}},\hat{\bm{\Xi}}^{*(j)}_{\tau^{*}_{p}})=\Biggl[
\textbf{g}_{1}(\bm{\Psi}^{(j-1)}_{\tau^{*}_{p}},\hat{\bm{\Xi}}^{*(j)}_{\tau^{*}_{p}})',\ldots, \textbf{g}_{r}(\bm{\Psi}^{(j-1)}_{\tau^{*}_{p}},\hat{\bm{\Xi}}_{\tau^{*}_{p}}^{*(j)})'\Biggr]'
\]
where
\[\textbf{g}_{k}(\bm{\Psi}^{(j-1)}_{\tau^{*}_{p}},\hat{\bm{\Xi}}^{*(j)}_{\tau^{*}_{p}})=\begin{pmatrix}
	\mathcal{N}\Bigg(\sum_{l=1}^{r}\sum_{\tau_{p}=1}^{\overline{\tau_{p}}}\hat{\lambda}^{*(j)}_{lk,\tau_{p}}\sum_{j=1}^{n}G_{1j,\tau_{p}}\psi_{jl,\tau_{p}}^{(j-1)}+\textbf{x}'_{1,\tau_{p}}\bm{\alpha}^{*(j)}_{k,\tau_{p}}+\bm{\iota}_{k,\tau_{p}}\bm{\gamma}^{*(j)}_{k,\tau_{p}}\Bigg)\\
	\vdots\\
	\mathcal{N}\Bigg(\sum_{l=1}^{r}\sum_{\tau_{p}=1}^{\overline{\tau_{p}}}\hat{\lambda}^{*(j)}_{lk,\tau_{p}}\sum_{j=1}^{n}G_{nj,\tau_{p}}\psi_{jl,\tau_{p}}^{(j-1)}+\textbf{x}'_{n,\tau_{p}}\bm{\alpha}^{*(j)}_{k,\tau_{p}}+\bm{\iota}_{k,\tau_{p}}\bm{\gamma}^{*(j)}_{k,\tau_{p}}\Bigg)
\end{pmatrix}\]
for $k=1,\ldots, r$. Update $\Psi^{(j-1)}_{\tau^{*}_{p}}$ in Step 1 to $\Psi^{j}_{\tau^{*}_{p}}$. Repeat Steps 1 and 2 until the process converges.

\cite{Pesendorfer2010} and \cite{Kasahara2012} shown that a key determinant of the convergence of the
NPL algorithm is the contraction property of the fixed point mapping \eqref{eq:eq}.
\section*{\large Conclusion}

In this paper, we develop a structural model of network formation. We characterize networks formation as a discrete game with incomplete information, where the decision of forming a link is characterized by a dyadic index such capture network externalities. We find that even with incomplete information and certain effort-dependent utilities, the Nash-Bayesian equilibrium is unique on community identification. This feature of the model means that the spillover effect is partially internalized in different quantities and consequently the Manski's effect can be controlled locally. An interesting extension of our work is to perform asymptotic dynamical inferences on the structural formation of communities. These extensions are left for future research.

\section*{\large Appendix}
\subsection*{\small A.1 Monte Carlo simulations}

In this section, we investigate the finite sample performance of procedures in Monte Carlo simulations. As in \cite{Graham2017}, we simulate networks using the family of rules
\[G_{ij}=\textbf{1}\{X_{i}X_{j}\beta +A_{i}+A_{j}-\nu_{ij}\geq 0\}\]
where $\delta=0$, $\beta=1$ and $X_{i}\in\{-1,1\}$, $i=1,\ldots, n$ are i.i.d binary variables with $\mathbb{P}(X_{i}=1)=1/2=\mathbb{P}(X_{i}=-1)$ and  simulated by
\[X_{i}=1-2\cdot\textbf{1}\{i\,\, \text{is even}\}\]
and $A_{i}=\Big(\dfrac{n-i}{n-1}\Big)C_{n}$,  where $C_{n}\in \{\log \log n, \log^{1/2}n, 2\log ^{1/2}n, \log n\}$ is a sparsity parameter.\footnote{\cite{Graham2017} modeled each $A_{i}=\alpha_{l}\textbf{1}\{X_{i}=-1\}+\alpha_{h}\textbf{1}\{X_{i}=1\}+V_{i}$, with $\alpha_{l}\leq \alpha_{h}$ and $V_{i}|X_{i}\sim \{Beta(\lambda_{0},\lambda_{1})-\dfrac{\lambda_{0}}{\lambda_{0}+\lambda_{1}}\}$.} With this specification, players with an even index prefer links
to agents with an even index over links to agents with an odd index, and vice versa for players with an odd index. The parameter of fixed effect specifications has first been proposed in \cite{Yan2016} and has also been used in \cite{Dzemski2019}. The goal is to recover the homophily coefficient. Let the density of a network be dened as the fraction of possible links that are observed, i.e., $\sum_{i=1}^{n}\sum_{j=1, j\neq i}^{n}\dfrac{G_{ij}}{n(n-1)}$. As in \cite{Dzemski2019}, the reciprocity parameter is set to $\rho=0, 0.5$. We consider two different network sizes: (i) $n=200$,  corresponding to $\binom{200}{2}=19,900$ dyads and (ii) $n=250$, corresponding to $\binom{250}{2}=31,125$ dyads. For each design and network size, we complete 700 Monte Carlo replications. All rejection probabilities are calculated based on a nominal level of $\alpha=0.1$.

Formally, each Monte Carlo designs satisfy the regularity conditions required for consistency and asymptotic normality $\hat{\beta}_{JML}$. However, in practice, the designs involve varying levels of link density.

Table 1 summarizes simulation results for the homophily and the reciprocity parameter. The maximum likelihood estimator $\hat{\beta}$ exhibits a bias of up to more than than one standard
deviation. The quality of the analytical bias correction decreases the sparser the design is. In the sparsest case, slightly less than half of the bias is eliminated. Without link reciprocity ($\hat{\rho}=0$), the maximum likelihood estimator $\hat{\rho}$ of the reciprocity is
approximately unbiased and analytical bias correction is not beneficial. With link reciprocity ($\hat{\rho}=0,5$),  exhibits a positive bias that is detected by the analytical bias correction. In the designs with extreme sparsity ($C_{n}=\log n$), the
$t$-test exhibits a size distortion. For one design with extreme sparsity ML estimation becomes numerically unstable.
\begin{table}[h!]\label{table1}
	\begin{center}\footnotesize
		\begin{tabular}{@{}ccrcccccr@{\hspace{0.000000em}} cc}
			\hline \hline\\
			&  &  &  & \multicolumn{4}{c}{Bias} &  & \multicolumn{2}{c}{Rej. Prob}\\ 
			\cmidrule{5-8}\cmidrule{10-11}$n$ & $\rho$ & $C_n$ & Density & $\hat{\beta}$ & $\hat{\beta}$ BC & $\hat{\rho}$ & $\hat{\rho}$ BC &  & $\hat{\beta}$ & $\hat{\rho}$
			\\ \midrule
			200 & 0.0 & $\log \log n$ & 0.15 & 1.18 & 0.11 & -0.01 & -0.03 &  & 0.05 & 0.08\\ 
			200 & 0.0 & $\log^{1/2} n$ & 0.09 & 1.07 & 0.18 & -0.00 & -0.05 &  & 0.07 & 0.07\\ 
			200 & 0.0 & $2\log^{1/2} n$ & 0.05 & 0.68 & 0.30 & -0.11 & -0.19 &  & 0.05 & 0.09\\ 
			200 & 0.0 & $\log n$ & 0.02 & 0.90 & 0.62 & -0.12 & -0.18 &  & 0.07 & 0.14\\ 
			200 & 0.5 & $\log \log n$ & 0.15 & 1.09& 0.11 & 0.17 & -0.17 &  & 0.12 & 0.09\\ 
			200 & 0.5 & $\log^{1/2} n$ & 0.09 & 0.60 & 0.19 & 0.46 & -0.30 &  & 0.12 & 0.09\\ 
			200 & 0.5 & $2\log^{1/2} n$ & 0.05 & 0.69 & 0.32 & 0.60 & 0.05 &  & 0.10 & 0.10\\ 
			200 & 0.5 & $\log n$ & 0.02 & - & - & - & - &  & - & -\\ 
			250 & 0.0 & $\log \log n$ & 0.15 & 0.80 & 0.01 & 0.03 & 0.03 &  & 0.9 & 0.05\\ 
			250 & 0.0 & $\log^{1/2} n$ & 0.09 & 1.10 & 0.14 & -0.14 & -0.09 &  & 0.14 & 0.10\\ 
			250 & 0.0 & $2\log^{1/2} n$ & 0.05 & 0.96 & 0.28 & -0.01 & -0.13 &  & 0.11 & 0.07\\ 
			250 & 0.0 & $\log n$ & 0.02 & 0.70 & 0.35 & -0.16 & -0.30 &  & 0.12 & 0.11\\ 
			250 & 0.5 & $\log \log n$ & 0.15 & 1.00 & 0.06 & 0.33 & -0.16 &  & 0.24 & 0.14\\ 
			250 & 0.5 & $\log^{1/2} n$ & 0.09 & 1.10 & 0.21 & 0.42 & -0.15 &  & 0.21 & 0.12\\ 
			250 & 0.5 & $2\log^{1/2} n$ & 0.05 & 1.04 & 0.26 & 0.64 & 0.17&  & 0.23 & 0.12\\ 
			250 & 0.5 & $\log n$ & 0.02 & 0.67 & 0.35 & 0.55 & 0.45 &  & 0.19 & 0.01\\ 
			
			\hline
		\end{tabular}
	\end{center}
	\caption{Simulation results fo $\hat{\beta}$ and $\hat{\rho}$. The bias is reported in term of standar deviations. $\hat{\beta}$ BC and $\hat{\rho}$ BC give results for the bias-corrected estimators $\hat{\beta}-W_{1,n}B^{\beta}_{n}/n$ and $\hat{\rho}-2(Q'_{n}W^{-1}_{1,n}B^{\rho}_{n}+B^{\rho}_{n})$ respectively. The empirical rejection probabilities ("Rej. Prob") are for two-sided t-test. Missing results (--) are reported if simulation runs are aborted due to numerical instability.}
\end{table}

\begin{table}[h!]\label{table2}
	\begin{center}\footnotesize
		\begin{tabular}{@{}ccrcrrr@{\hspace{0.000000em}} rrccc}
			\hline \hline\\
			&  &  &  & \multicolumn{2}{c}{In-degree} &  & \multicolumn{2}{c}{Out-degree} \\ 
			\cmidrule(lr){5-6} \cmidrule(lr){7-9} $n$ & $\rho$ & $C_n$ & Density & Mean & Median &  & Mean & Median & Comp. Connected & Min. cut & Clustering
			\\ \midrule
			200 & 0.0 & $\log \log n$ & 0.15 & 30.25 & 28.01 &  & 30.25 & 28.04 & 1.00 & 4.10 & 0.53\\ 
			200 & 0.0 & $\log^{1/2} n$ & 0.09 & 18.00 & 14.35 &  & 18.00 & 14.33 & 0.98 & 0.02 & 0.46\\ 
			200 & 0.0 & $2\log^{1/2} n$ & 0.05 & 9.33 & 4.99 &  & 9.33 & 4.96 & 0.74 & 0.00 & 0.44\\ 
			200 & 0.0 & $\log n$ & 0.02 & 3.53 & 0.02 &  & 3.53 & 0.02 & 0.41 & 0.00 & 0.43\\ 
			200 & 0.5 & $\log \log n$ & 0.15 & 30.24 & 27.97 &  & 30.24 & 28.00 & 1.00 & 4.12 & 0.47\\ 
			200 & 0.5 & $\log^{1/2} n$ & 0.09 & 17.99 & 14.32 &  & 17.99 & 14.30 & 0.98 & 0.02 & 0.42\\ 
			200 & 0.5 & $2\log^{1/2} n$ & 0.05 & 9.33 & 5.00 &  & 9.33 & 4.98 & 0.74 & 0.00 & 0.39\\ 
			200 & 0.5 & $\log n$ & 0.02 & 3.53 & 0.02 &  & 3.53 & 0.02 & 0.42 & 0.00 & 0.39\\ 
			250 & 0.0 & $\log \log n$ & 0.15 & 36.58 & 33.62 &  & 36.58 & 33.60 & 1.00 & 4.92 & 0.52\\ 
			250 & 0.0 & $\log^{1/2} n$ & 0.09 & 21.72 & 16.99 &  & 21.72 & 16.99 & 0.99 & 0.03 & 0.46\\ 
			250 & 0.0 & $2\log^{1/2} n$ & 0.05 & 11.21 & 5.76 &  & 11.21 & 5.78 & 0.75 & 0.00 & 0.44\\ 
			250 & 0.0 & $\log n$ & 0.02 & 4.07 & 0.00 &  & 4.07 & 0.01 & 0.41 & 0.00 & 0.43\\ 
			250 & 0.5 & $\log \log n$ & 0.15 & 36.59 & 33.59 &  & 36.59 & 33.63 & 1.00 & 4.93 & 0.47\\ 
			250 & 0.5 & $\log^{1/2} n$ & 0.09 & 21.74 & 17.03 &  & 21.74 & 16.98 & 0.99 & 0.02 & 0.42\\ 
			250 & 0.5 & $2\log^{1/2} n$ & 0.05 & 11.22 & 5.77 &  & 11.22 & 5.76 & 0.75 & 0.00 & 0.40\\ 
			250 & 0.5 & $\log n$ & 0.02 & 4.07 & 0.00 &  & 4.07 & 0.01 & 0.42 & 0.00 & 0.39\\ 
			\hline 
		\end{tabular}
	\end{center}
	\caption{"In-degree", and "Out-degree" gives average network summary statistics across the 700 Monte Carlo repetitions for each design. Across all designs $X_{i}\in \{-1,1\}$ with $\mathbb{P}(X_{i})=\mathbb{P}(X_{i}=-1)=1/2$ and $\beta=1$. Summary network statistics are presented for each design. "Component connected"=share of players belonging to the largest connected component, "Min. cut"= minimum cut of the network, "Clustering"= clustering coeficient.}
\end{table}

\subsection*{\small A.2. Proof of Proposition 1}

As $\textbf{g}(\cdot)$ is continuously differentiable then:
$\|\textbf{g}(\bm{\psi})-\textbf{g}(\bm{\xi})\|_{1,\tau_{p}}\leq \Bigg[\sup_{\alpha\in[\xi_{i},\psi_{i}]}\|\textbf{g}'(\alpha)\|_{1}\Bigg]\|\bm{\xi}-\bm{\psi}\|_{\tau_{p}}$ for alll $i$ with
\[\dfrac{\partial \textbf{g}(\bm{\psi})}{\partial \bm{\psi_{\tau_{p}}}'}=\begin{pmatrix}
	\dfrac{\partial \textbf{g}_{1}(\bm{\psi})}{\partial \bm{\psi}'_{1}}&\cdots &\dfrac{\partial \textbf{g}_{1}(\bm{\psi})}{\partial \bm{\psi}'_{r}}\\
	\vdots & &\vdots\\
	\dfrac{\partial \textbf{g}_{r}(\bm{\psi})}{\partial \bm{\psi}'_{1}}&\cdots &\dfrac{\partial \textbf{g}_{r}(\bm{\psi})}{\partial \bm{\psi}'_{r}}
\end{pmatrix}\]
where
\[\dfrac{\partial g_{k}(\bm{\psi})}{\partial \bm{\psi}'_{l}}=\tilde{\lambda}_{lk,\tau_{p}}\begin{pmatrix}
	G_{11,\tau_{p}}f_{k,\tau_{p}}(q_{ik})&\cdots& G_{1n,\tau_{p}}f_{k}(q_{1k})\\
	\vdots&\vdots&\vdots\\
	G_{n1,\tau_{p}}f_{k,\tau_{p}}(q_{nk})&\cdots&G_{nn,\tau_{p}}f_{k,\tau_{p}}(q_{nk})
\end{pmatrix}\]
It follows that Assumption 4
\[\sup_{\alpha\in[\xi_{i},\psi_{i}]}\|\textbf{g}'(\alpha)\|_{1}\leq \max_{k=1,\ldots,r}|\tilde{\lambda}_{lk,\tau_{p}}|\max_{i=1,\ldots, r}\sum_{j=1}^{n}G_{ij,\tau_{p}}\max_{k}\sup_{q}f_{k}(q)=\|\tilde{\bm{\Lambda}}_{\tau_{p}}\|_{1}\max_{k}\sup_{q}f_{k}(q)\quad \forall i\]
Analogously for $\sup_{\alpha\in[\xi_{i},\psi_{i}]}\|\textbf{g}'(\alpha)\|_{\infty}$. Then by Banach Fixed Point Theorem $\textbf{g}(\cdot)$ has a unique solution $\bm{\psi}_{\tau_{p}}$ such that $\bm{\psi}=g(\bm{\psi})$.

\subsection*{\small A.2. Proof of Proposition 2}

According Theorem 1 in \cite{Jackson2001}, if there is a function $\Theta:\mathcal{G}\to \mathbb{R}$ such that for any networks $G$, $G'$ that differ by one link, $G'$ defeats $G$ if and only if $\Theta(G')>\Theta(G)$, then there is no closed cycle. In the case of TU $G'$ defeating $G$ means that for any $i\neq j$ such that $G'_{ij}\neq G_{ij}$, $U_{i}(G')+U_{j}(G')>U_{i}(G)+U_{j}(G)$. Hence, the proof is complete if we can find $\Theta$ for the utility function \eqref{eq:Utility}.

We show that 
\[\Theta(G)=\sum_{i=1}^{n}\Bigg(\sum_{k=1}^{r}\Big(\sum_{l=1}^{r}s_{lk}\Big)\sum_{j=1}^{n}G_{ij}a_{jl}a_{ik}+w_{ik}-\epsilon_{ik}\Bigg)y_{ik}\]
has the desired property. Assume without loss of generality that $G=(0, G_{-ij})$ and $G'=(1,G_{-ij})$. It suffices
to show that $\Theta(G')-\Theta(G)=\Delta \mathbb{E}(U_{ij}(G_{-ij}))+\Delta \mathbb{E}(U_{ji}(G_{-ji}))$. A simple calculation shows that:
\[\Theta(G')-\Theta(G)=\sum_{k=1}^{r}\Big(\sum_{l=1}^{r}s_{lk}a_{jl}a_{ik}\Big)y_{ik}+\sum_{k=1}^{r}\Big(\sum_{l=1}^{r}s_{lk}a_{il}a_{jk}\Big)y_{jk}\]
While, from \eqref{eq:esp} we have

\begin{eqnarray}
	\Delta \mathbb{E}(U_{ij}(G_{-ij},y_{ik}))=\sum_{k=1}^{r}\big(\sum_{l=1}^{r}s_{lk}\psi_{jl}\big)y_{ik}\\
	\Delta \mathbb{E}(U_{ji}(G_{-ji},y_{jk}))=\sum_{k=1}^{r}\big(\sum_{l=1}^{r}s_{lk}\psi_{il}\big)y_{jk}
\end{eqnarray}

$\Theta(G')-\Theta(G)=\Delta \mathbb{E}(U_{ij}(G_{-ij},y_{ik}))+\Delta \mathbb{E}(U_{ji}(G_{-ij},y_{ik}))$. The proof is complete.

\subsection*{\small A.3. Proof of Proposition 3}
	
Note that
\[\begin{split}
	L(\theta,\textbf{A}_{n})=&\sum_{i\neq j}\Biggl[G_{ij}\ln p_{ij}(\theta,\textbf{A})+(1-G_{ij})\ln \Big(1-p_{ij}(\theta,\textbf{A})\Big)\Biggr]\\
	=&\sum_{i\neq j}\Biggl[G_{ij}\ln\big(\dfrac{p_{ij}(\theta,A_{i},A_{j})}{1-p_{ij}(\theta,A_{i},A_{j})}\big)+\ln\Big(1-p_{ij}(\theta,A_{i},A_{j})\Big)\Biggr]\\
	=&\sum_{i\neq j}\Biggl[G_{ij}\ln\bigg(\dfrac{p_{ij}(\theta,A_{i},A_{j})}{1-p_{ij}(\theta,A_{i},A_{j})}\bigg)-p_{ij}\ln\bigg(\dfrac{p_{ij}(\theta,A_{i},A_{j})}{1-p_{ij}(\theta,A_{i},A_{j})}\bigg)-p_{ij}\ln\bigg(\dfrac{p_{ij}}{p_{ij}(\theta,A_{i},A_{j})}\bigg)-\mathbb{H}(p_{ij})\Biggr]\\
	=&\sum_{i\neq j}(G_{ij}-p_{ij})\ln\bigg(\dfrac{p_{ij}(\theta,A_{i},A_{j})}{1-p_{ij}(\theta,A_{i},A_{j})}\bigg)-\sum_{i\neq j}D_{KL}(p_{ij}||p_{ij}(\theta,A_{i},A_{j}))-\sum_{i\neq j}\mathbb{H}(p_{ij})
\end{split}\]
for $D_{KL}(p_{ij}||p_{ij}(\theta,A_{i},A_{j}))$ the Kullback–Leibler divergence of $p_{ij}(\theta,A_{i},A_{j})$ from $p_{ij}$ and $\mathbb{H}(p_{ij})$ the binary entropy function. For all $(\beta,\delta)\in \mathbb{A}\times \mathbb{B}$, $\textbf{A}\in \mathbb{I}^{n}$, and $\textbf{X}\in \mathbb{X}^{n}$,
\[\bigg|\binom{n}{2}^{-1}\sum_{i=1}^{n}\sum_{i\neq j}(G_{ij}-p_{ij})\ln\bigg(\dfrac{p_{ij}(\theta,A_{i},A_{j})}{1-p_{ij}(\theta,A_{i},A_{j})}\bigg)\bigg|\leq \dfrac{2}{n}\sum_{i=1}^{n}\dfrac{1}{n-1}\sum_{i\neq j}(G_{ij}-p_{ij})\ln \bigg(\dfrac{p_{ij}(\theta,A_{i},A_{j})}{1-p_{ij}(\theta,A_{i},A_{j})}\bigg) \]
We can apply a Hoeffding inequality to the terms in the outer summand to the right of the inequality above.	
Let $\vartheta_{ij}(\theta,A_{i},A_{j})=\ln \bigg(\dfrac{p_{ij}(\theta,A_{i},A_{j})}{1-p_{ij}(\theta,A_{i},A_{j})}\bigg)$ and $\varphi=\ln \dfrac{1-\xi}{\xi}$. Condition \eqref{eq:p_ij} implies that that -$\varphi\leq \varphi_{lk}(\theta,A_{i},A_{j})\leq \varphi$. Hoeffding’s inequality therefore gives
\[\mathbb{P}\bigg(\bigg|\dfrac{1}{n-1}\sum_{i\neq j}(G_{ij}-p_{ij})\varphi_{ij}(\theta,A_{i},A_{j})\bigg|\geq \nu\bigg)\leq 2\exp\bigg(-\dfrac{(n-1)\nu^{2}}{2(1-\xi)^{2}\varphi^{2}}\bigg)\]
From Lemma 3 in \cite{Graham2017} implies that, with probability equall to $1-O(n^{-2})$, and for any $\theta\in\mathbb{A}\times \mathbb{B}$, $\textbf{A}\in \mathbb{I}^{n}$,
\[\bigg|\binom{n}{2}^{-1}\sum_{i=1}^{n}\sum_{i\neq j}(G_{ij}-p_{ij})\ln\bigg(\dfrac{p_{ij}(\theta,A_{i},A_{j})}{1-p_{ij}(\theta,A_{i},A_{j})}\bigg)\bigg|< O\bigg(\sqrt{\dfrac{\ln n}{n}}\bigg)\]
and hence that
\begin{equation}\label{eq:sup}\sup_{(\beta,\delta)\in\mathbb{A}\times \mathbb{B}, \textbf{A}\in \mathbb{I}^{n}}\bigg|\binom{n}{2}^{-1}\sum_{i=1}^{n}\sum_{i\neq j}(G_{ij}-p_{ij})\ln\bigg(\dfrac{p_{ij}(\theta,A_{i},A_{j})}{1-p_{ij}(\theta,A_{i},A_{j})}\bigg)\bigg|< O\bigg(\sqrt{\dfrac{\ln n}{n}}\bigg)
\end{equation}
From $L(\theta,\textbf{A}_{n})$ and \eqref{eq:sup} therefore give, again with probability equal to $1-O(n^{-2})$, the uniform convergence result
\begin{equation}\label{eq:sup_1}
	\sup_{(\beta,\delta)\in\mathbb{A}\times \mathbb{B}, \textbf{A}\in \mathbb{I}^{n}}\bigg|\binom{n}{2}^{-1}\biggl\{L(\theta,\textbf{A})-\mathbb{E}[L(\theta,\textbf{A}|\textbf{X},\textbf{A}_{0})]\biggr\}]\bigg|< O\bigg(\sqrt{\dfrac{\ln n}{n}}\bigg)
\end{equation}
Let $A_{0}\times B_{0}$ be an open neighborhood in $\mathbb{A}\times \mathbb{B}$ with product topology which contains $(\beta_{0},\delta_{0})$ and $(A_{0}\times B_{0})^{c}$ its complement in $\mathbb{A}\times \mathbb{B}$.

Define
\[\kappa_{n}=\max_{\textbf{A}\in \mathbb{I}^{n}} \binom{n}{2}^{-1}L(\theta,\textbf{A}|\textbf{X},\textbf{A}_{0})-\max_{(\beta,\delta)\in (A_{0}\times B_{0})^{c}, \textbf{A}\in \mathbb{I}^{n}}\binom{n}{2}^{-1}\mathbb{E}[L(\theta,\textbf{A}|\textbf{X},\textbf{A}_{0})]\]
As long as $\mathbb{E}[L_{n}(\theta,\textbf{A}|\textbf{X},\textbf{A}_{0})]$ is uniquely maximized at $\beta_{0}$, $\delta_{0}$ and $\textbf{A}_{0}$, then $\kappa_{n}$ wil be strictly greather than zero by Assumption 6. Let $Z_{n}$ be the event
\[\bigg|\max_{\textbf{A}\in \mathbb{I}^{n}} \binom{n}{2}^{-1}\mathbb{E}[L(\theta,\textbf{A}|\textbf{X},\textbf{A}_{0})]-\max_{(\beta,\delta)\in (A_{0}\times B_{0})^{c}, \textbf{A}\in \mathbb{I}^{n}}\binom{n}{2}^{-1}\mathbb{E}[L(\theta,\textbf{A}|\textbf{X},\textbf{A}_{0})]\bigg|<\kappa_{n}/2\]
for all $\theta=(\beta,\delta)\in \mathbb{A}\times \mathbb{B}$. Under event $Z_{n}$, we get the inequalities
\begin{equation}\label{eq:b1}
	\max_{\textbf{A}\in \mathbb{I}^{n}}\binom{n}{2}^{-1}\mathbb{E}[L(\tilde{\theta},\textbf{A}|\textbf{X},\textbf{A}_{0})]>\binom{n}{2}^{-1}L(\tilde{\theta},\tilde{\textbf{A}}|\textbf{X},\textbf{A}_{0})-\kappa_{n}/2
\end{equation}
and 
\begin{equation}\label{eq:b2}
	\max_{\textbf{A}\in \mathbb{I}^{n}}\binom{n}{2}^{-1}L(\theta_{0},\textbf{A}|\textbf{X},\textbf{A}_{0})>\max_{\textbf{A}\in \mathbb{I}^{n}}\binom{n}{2}^{-1}\mathbb{E}[L(\theta_{0},\tilde{\textbf{A}}|\textbf{X},\textbf{A}_{0})]-\kappa_{n}/2
\end{equation}
By definition of the MLE we have tha $\binom{n}{2}^{-1}L(\tilde{\theta},\tilde{\textbf{A}})\geq \max_{\textbf{A}\in \mathbb{I}^{n}}\binom{n}{2}^{-1}L(\theta_{0},\textbf{A}) $ and, making of use \eqref{eq:b1} 
\begin{equation}\label{eq:b3}
	\max_{\textbf{A}\in \mathbb{I}^{n}}\binom{n}{2}^{-1}\mathbb{E}[L(\tilde{\theta},\textbf{A}|\textbf{X},\textbf{A}_{0})]>\max_{\textbf{A}\in \mathbb{I}^{n}}\binom{n}{2}^{-1}L(\theta_{0},\textbf{A})-\kappa_{n}/2
\end{equation}
From \eqref{eq:b2} and \eqref{eq:b3}
\begin{align}\label{eq:b4}
	\max_{\textbf{A}\in \mathbb{I}^{n}}\binom{n}{2}^{-1}\mathbb{E}[L(\tilde{\theta},\textbf{A}|\textbf{X},\textbf{A}_{0})]>&\max_{\textbf{A}\in \mathbb{I}^{n}}\binom{n}{2}^{-1}\mathbb{E}[L(\theta_{0},\textbf{A}|\textbf{X},\textbf{A}_{0})]-\kappa_{n}\\
	=&\max_{(\beta,\delta)\in (A_{0}\times B_{0})^{c},\textbf{A}\in \mathbb{I}^{n}}\binom{n}{2}^{-1}\mathbb{E}[L(\theta,\textbf{A}_{0}|\textbf{X},\textbf{A}_{0})]
\end{align}
from \eqref{eq:b4}, we have that $Z_{n}$ implies $\tilde{\theta}\in A_{0}\times B_{0}$. Therefore, $\mathbb{P}(Z_{n})\leq \mathbb{P}(\tilde{\theta}\in A_{0}\times B_{0})$, But \eqref{eq:sup_1} imlies that $\lim_{n\to \infty}\mathbb{P}(Z_{n})=1$ and hence $\tilde{\theta}\xrightarrow{p} \theta_{0}$.

\subsection*{\small A.4. Proof of Lemma 1}

For a random variable $R$ we set $\overline{R}=\mathbb{E}(R)$. Omitting function argument indicates that the function is eliminated at the true paramters,e.g., $\textbf{H}=\textbf{H}(\theta_{0},\textbf{A}_{0})$ In particular, for $b>0$
\[\overline{\textbf{H}}=\overline{\textbf{H}}^{*}+b(\textbf{u}_{n}\textbf{u}'_{n})/n\]
where $\textbf{u}_{n}=(\iota_{n},-\iota_{n})'$,
\[\textbf{H}^{*}=\begin{pmatrix}
	\overline{\textbf{H}}^{*}_{11}&\overline{\textbf{H}}^{*}_{12}\\
	\overline{\textbf{H}}^{*}_{21}&\overline{\textbf{H}}^{*}_{22}
\end{pmatrix}\]
with $\overline{\textbf{H}}^{*}_{11}=\diag\Big(\big(\dfrac{1}{n}\sum_{j=1,j\neq i}\rho_{ij}\big)_{i=1}^{n}\Big)$, $\overline{\textbf{H}}^{*}_{22}=\diag\Big(\big(\dfrac{1}{n}\sum_{i=1,i\neq j}\rho_{ij}\big)_{j=1}^{n}\Big)$ and $\overline{\textbf{H}}^{*}_{12}$ is a $n\times n$ matrix with diagonal entries equal to zero and off-diagonal entries $\rho_{ij}/n$ for $i,j\in[n]$ and $i\neq j$.

 We have
\begin{equation}
	\begin{split}
	\partial_{\varrho}M=&\dfrac{1}{n}\sum_{i=1}^{n}\sum_{j=1,i<j}^{n}G_{ij}G_{ji}\Big[\dfrac{\partial_{\varrho}(s_{ij})}{s_{ij}}+\dfrac{\partial_{\varrho}s_{ij}}{1-s_{ij}}\Big]-\dfrac{\partial_{\rho}s_{ij}}{1-s_{ij}}\\
	=&\dfrac{1}{n-1}\sum_{i=1}^{n}\sum_{j=1,i<j}^{n}J_{ij}(G_{ij}G_{ji}-s_{ij}).
\end{split}
\end{equation}
Then
\[\partial_{\varrho^{2}}M=\dfrac{1}{n}\sum_{i=1}^{n}\sum_{j=1,i<j}^{n}\Big(\partial_{\varrho}J_{ij}(G_{ij}G_{ji}-s_{ij})+J_{ij}(-\partial_{\rho}s_{ij})\Big)\]
Note that
\[\mathbb{E}\Biggl[\Bigg(\dfrac{1}{n-1}\sum_{i=1}^{n}\sum_{j=1,i<j}^{n}\partial_{\varrho}J_{ij}(G_{ij}G_{ji}-s_{ij})\Bigg)^{2}\Biggr]=O_{p}(1)\]
and therefore
\[\partial_{\varrho^{2}}M=-\dfrac{1}{n-1}\sum_{i=1}^{n}\sum_{j=1,i<j}^{n}J_{ij}(\partial_{\rho}s_{ij})+O_{p}(1)\]
Arguing similarly we get
\[\begin{split}
	\partial_{\varrho \delta}M=&-\dfrac{1}{n}\sum_{i=1}^{n}\sum_{j=1,i<j}^{n}J_{ij}(\partial_{\delta}s_{ij})+O_{p}(1)\\
	=&-\dfrac{1}{n}\sum_{i=1}^{n}\sum_{j=1,i<j}^{n}J_{ij}\Big((\partial_{g_{1}}s_{ij})(\sum_{k\in J}G_{ij}G_{ik}G_{jk}+(\partial_{g_{2}}s_{ij})(\sum_{k\in J}G_{ij}G_{ik}G_{jk})\Big)+O_{p}(1)\\
	=&-\dfrac{1}{n}\sum_{i=1}^{n}\sum_{j=1,i<j}^{n}\Big(\sum_{k\in J}G_{ij}G_{ik}G_{jk}J_{ij}(\partial_{g_{1}}s_{ij}+\partial_{g_{2}}s_{ij})\Big)+O_{p}(1)
\end{split}\]
Also
\[\begin{split}
		\partial_{\varrho \beta}M=&-\dfrac{1}{n}\sum_{i=1}^{n}\sum_{j=1,i<j}^{n}J_{ij}(\partial_{\beta}s_{ij})+O_{p}(1)\\
		=&-\dfrac{1}{n}\sum_{i=1}^{n}\sum_{j=1,i<j}^{n}J_{ij}(\partial_{g_{1}}s_{ij}X'_{ij}+\partial_{g_{2}}s_{ij}X'_{ji})+O_{p}(1)\\
		=&-\dfrac{1}{n}\sum_{i=1}^{n}\sum_{j=1,i<j}^{n}J_{ij}X'_{ij}(\partial_{g_{1}}s_{ij}+\partial_{g_{2}}s_{ij})+O_{p}(1)\\
		=&-\dfrac{1}{n}\sum_{i=1}^{n}\sum_{j=1,j\neq i}^{n}J_{ij}X'_{ij}(\partial_{g_{1}}s_{ij})+O_{p}(1)
\end{split}\]
For $k=1,\ldots, \dim(\beta)$ let
\[\nabla_{ij,k}=-\dfrac{1}{n}\sum_{k_{1}=1}^{n}\sum_{k_{2}=1,k_{2}\neq k_{1}}^{n}\Bigg(H^{-1}_{11,ik_{1}}+H^{-1}_{12,jk_{1}}+H^{-1}_{21,ik_{2}}+H^{-1}_{22,jk_{2}}\Bigg)\mathbb{E}[\partial_{\theta_{k}}l_{k_{1}k_{2}}]\]
where $l_{k_{1}k_{2}}=G_{k_{1}k_{2}}\ln(p_{k_{1}k_{2}})+(1-G_{k_{1}k_{2}})\ln(1-p_{k_{1}k_{2}})$ and let $\nabla_{ij}=(\nabla_{ij,1},\ldots, \nabla_{ij,\dim(\beta)})'$. By Lemma S.8 in \cite{Fernandez2016} and Lemma C.11 in \cite{Dzemski2019} 
\[(\partial_{\varrho A}M)(H^{-1})(\partial_{\varrho \beta'}L)=\dfrac{1}{n}\sum_{i=1}^{n}\sum_{j=1,\neq i}^{n}J_{ij}s_{ij}\nabla'_{ij}+O_{p}(1)\]

\subsection*{\small A.5. Proof of Lemma 2}

By definition
\[n^{-2}\Big[\widehat{\mathbb{E}[S_{n}]}-S_{n}\Big]=s_{n}(\hat{\theta},\hat{\textbf{A}})-s_{n}(\theta_{0},\textbf{A}_{0})\]
From Taylor expansion
\begin{multline*}
	s_{n}(\hat{\theta},\hat{\textbf{A}})-s_{n}(\theta_{0},\textbf{A}_{0})=\partial_{\beta}s_{n}(\theta_{0},\textbf{A}_{0})(\beta_{JML}-\beta_{0})+\partial_{\delta}s_{n}(\theta_{0},\textbf{A}_{0})(\delta_{JML}-\delta_{0})+\partial_{\textbf{A}}s_{n}(\theta_{0},\textbf{A}_{0})(\hat{\textbf{A}}-\textbf{A}_{0})+\\
	+\dfrac{1}{2}(\hat{\textbf{A}}-\textbf{A}_{0})\Big(\partial_{\textbf{A}\textbf{A}'}s_{n}(\theta_{0},\textbf{A}_{0})(\hat{\textbf{A}}-\textbf{A}_{0})\Big)+\dfrac{1}{2}(\delta_{JML}-\delta_{0})\Big(\partial_{\delta \delta'}s_{n}(\theta_{0},\textbf{A}_{0})(\delta_{JML}-\delta_{0})\Big)+\\
	+\dfrac{1}{2}(\beta_{JML}-\beta_{0})\Big(\partial_{\beta \beta'}s_{n}(\theta_{0},\textbf{A}_{0})(\beta_{JML}-\beta_{0})\Big)\\
	+\dfrac{1}{2}(\delta_{JML}-\delta_{0})\Big(\partial_{\beta\delta'}s_{n}(\theta_{0},\textbf{A}_{0})(\beta_{JML}-\beta_{0})\Big)+\dfrac{1}{2}(\delta_{JML}-\delta_{0})\Big(\partial_{\textbf{A}\delta'}s_{n}(\theta_{0},\textbf{A}_{0})(\hat{\textbf{A}}-\textbf{A}_{0})\Big)+\\
	+\dfrac{1}{2}(\beta_{JML}-\beta_{0})\Big(\partial_{\textbf{A}\beta'}s_{n}(\theta_{0},\textbf{A}_{0})(\hat{\textbf{A}}-\textbf{A}_{0})\Big)+\dfrac{1}{6}\sum_{h=1}^{\dim(\textbf{A})}(\hat{\textbf{A}}-\textbf{A}_{0})'[\partial_{\textbf{A} \textbf{A}'\textbf{A}_{h}}s_{n}(\textbf{A}^{*},\theta_{0})](\hat{\textbf{A}}-\textbf{A}_{0})[\hat{\textbf{A}}-\textbf{A}_{0}]_{h}
\end{multline*}
Note that
\begin{multline*}
	n^{1-1/q}\Big|\Big|\partial_{\beta \textbf{A}}s_{n}(\theta_{0},\textbf{A}^{*})\Big|\Big|_{q}\Big|\Big|\hat{\textbf{A}}-\textbf{A}_{0}\Big|\Big|_{q}\Big|\Big|\beta_{JML}-\beta_{0}\Big|\Big|_{2}+\Big|\Big|\partial_{\beta \beta}s_{n}(\theta^{*}_{0},\textbf{A}_{0})\Big|\Big|_{q}\Big|\Big|\beta_{JML}-\beta_{0}\Big|\Big|_{q}\Big|\Big|\beta_{JML}-\beta_{0}\Big|\Big|^{2}_{2}+\\
	+\dfrac{1}{2}\Big|\Big|\partial_{\delta\delta' \delta}s_{n}(\delta^{*}_{0},\textbf{A}_{0})\Big|\Big|_{q}\Big|\Big|\delta_{JML}-\delta_{0}\Big|\Big|_{q}\Big|\Big|\delta_{JML}-\delta_{0}\Big|\Big|^{3}_{2}=O_{p}(n^{-1/2}+1/q)=o_{p}(1)
\end{multline*}
the result is obtained by evaluating $s$ in the true parameters.
\subsection*{\small A.6. Proof of Proposition 4}

We write
\[n^{-2}\Bigg[S_{n}-\widehat{\mathbb{E}[S_{n}]}\Bigg]=n^{-2}\Bigg[S_{n}-\mathbb{E}[S_{n}]\Bigg]-n^{-2}\Bigg[\widehat{\mathbb{E}[S_{n}]}-\mathbb{E}[S_{n}]\Bigg]\]
Fron Lemma 3 \begin{multline*}
	n^{-2}\Bigg[S_{n}-\widehat{\mathbb{E}[S_{n}]}\Bigg]=s_{n}(\hat{\theta},\hat{\textbf{A}})-s_{n}(\theta_{0},\textbf{A}_{0})=\Bigg[(\partial_{\beta'}s_{n})+(\partial_{\delta'}s_{n})+(\partial_{\textbf{A}}s_{n}H^{-1}(\partial_{\textbf{A}\beta'\delta})L)\Bigg](\hat{\theta}-\theta_{0})+\\
	\partial_{\textbf{A}}s_{n}H^{-1}S+B_{1,n}+o(1)
\end{multline*}
Also
\[\partial_{\beta} s_{n}=\dfrac{1}{n-1}\sum_{i=1}^{n}\sum_{j=1,j\neq i}^{n}\chi_{ij,n}\rho_{ij}X_{ij}\quad \partial_{\delta'}s_{n}=\dfrac{1}{n-2}\sum_{i=1}^{n}\sum_{j=1, j\neq i}^{n}\sum_{k=1, k\neq i,j}^{n}\bm{\chi}_{ij,n}\bm{\chi}_{jk,n}\rho_{ij}\rho_{jk}X_{ij}X_{jk}\]
We apply the Lemma 2 to
\[\nabla_{ij,k}=-\dfrac{1}{n}\sum_{k_{1}=1}^{n}\sum_{k_{2}=1,k_{2}\neq k_{1}}^{n}\Big(H^{-1}_{11,ik_{1}}+H^{-1}_{12,jk_{1}}+H^{-1}_{21,ik_{2}}+H^{-1}_{22,jk_{2}}\Big)\mathbb{E}[\partial_{\beta_{k}}l_{k_{1}}l_{k_{2}}]\]
By Lemma S.8 in \cite{Fernandez2016} and Lemma C.6 in \cite{Dzemski2019}
\begin{multline*}
	n^{-2}\Bigg[S_{n}-\widehat{\mathbb{E}[S_{n}]}\Bigg]=-B_{1,n}-V'_{n}W^{-1}_{1,n}B_{\beta,n}+V'_{n}W^{-1}_{1,n}B_{\delta,n}+\\
	\dfrac{1}{n-1}\sum_{i=1}^{n}\sum_{j=1,j\neq i}^{n}(\chi_{ij,n}-(\mathcal{P}\chi_{n})_{ij}-v_{ij,n})H_{ij}(G_{ij}-p_{ij})+o_{p}(1).
\end{multline*}

 \nocite{}
 \bibliographystyle{apalike}
\bibliography{Bliblio_2022_03}

\end{document}